\documentclass[12pt]{JHEP3}
\pdfoutput=1

\usepackage{amssymb, amsmath, amsopn, amsthm}
\usepackage{epsfig,amsfonts,bm}
\usepackage{graphics}

\def\bec{\begin{center}}
\def\eec{\end{center}}

\def\p{\partial}

\newcommand{\be}{\begin{equation}}
\newcommand{\ee}{\end{equation}}
\newcommand{\bea}{\begin{eqnarray}}
\newcommand{\eea}{\end{eqnarray}}
\newcommand{\beas}{\begin{eqnarray*}}
\newcommand{\eeas}{\end{eqnarray*}}

\def\scriptlap{{\kern1pt\vbox{\hrule height 0.8pt\hbox{\vrule width 0.8pt
  \hskip2pt\vbox{\vskip 4pt}\hskip 2pt\vrule width 0.4pt}\hrule height 0.4pt}
  \kern1pt}}

\def\Biggg#1{{\hbox{$\left#1\vbox to 25pt{}\right.\n@space$}}}
\def\n@space{\nulldelimiterspace=0pt \m@th}
\def\m@th{\mathsurround = 0pt}

\title{Study of Anisotropic Black Branes in Asymptotically anti-de Sitter}

\author{Norihiro Iizuka$^{1}$ and Kengo Maeda${}^2$
\\

{\small \sl ${}^1$Theory Division, CERN, CH-1211 Geneva 23, Switzerland} \\
{\small \tt  {norihiro.iizuka@cern.ch}} \\

{\small \sl ${}^2$Faculty of Engineering,
Shibaura Institute of Technology,} \\ 
{\small \sl Saitama, 330-8570, Japan} \\
{\small \tt {maeda302@sic.shibaura-it.ac.jp}}

\vspace{0.1cm}

}

\abstract{%
We investigate the four dimensional gravitational theories which admit homogeneous but 
anisotropic black brane solutions in asymptotically AdS space-time. 
The gravitational theories we consider are 1) Einstein-Maxwell-dilaton theory, and 2)  
Einstein-Maxwell-dilaton-axion theory with $SL(2,R)$ symmetry. 
We obtain the solutions both analytically and numerically.  
Analytical solutions are obtained by perturbation from the isotropic solutions.  
Our solutions approach singular behavior at the horizon in 
the extremal limit but in non-extremal case, they 
are smooth everywhere. We also discuss how the third law of thermodynamics holds in our set-up.

}

\preprint{CERN-PH-TH-2012-073}
\begin{document}

\section{Introduction}\label{sec:intro}

Understanding the strongly coupled limit of quantum field theory is 
a long standing 
problem in theoretical physics, 
which exists in broad range of physics, from nuclear physics to condensed matter theory. 
The AdS/CFT correspondence, or more broadly, the gauge/gravity correspondence 
\cite{Maldacena:1997re,Witten:1998qj,Gubser:1998bc}
is an extremely useful tool for this purpose  
since it gives a new perspective for strong coupled field theory from totally different theory viewpoint, 
 {\it i.e.,} gravitational theory on the asymptotically anti-de Sitter~(AdS) viewpoint. 
Given this, it is very interesting to search for the generic gravitational solutions 
which are asymptotically AdS at UV but show non-trivial behavior at the interior, especially in IR. 
Each of these generic geometry corresponds to interesting phase in the field theory side.  

Famous such examples are black brane solutions in asymptotically AdS space-time. 
These solutions are very useful gravitational backgrounds for studying  
the strongly coupled dual field theory, which 
corresponds to deconfined phase in QCD \cite{Witten:1998zw}. These black brane solutions, 
especially their charged ones, are also
useful to study    
superfluid phase \cite{Gubser:2008px,Hartnoll:2008vx,Hartnoll:2008kx} and ``fractionallized'' Fermi-liquid phase in condensed matter system \cite{Sachdev:2010um},
and they have been studied in great detail for the application to both QCD and 
condensed matter physics. For recent review of these, see for examples, 
\cite{SachdevMueller,Hartnoll:2009sz,Herzog:2009xv,McGreevy:2009xe,Horowitz:2010gk,Sachdev:2010ch, CasalderreySolana:2011us,Hartnoll:2011fn}.

Generically, finding generic black brane solutions of Einstein equations 
are quite difficult task, therefore we usually put several ansatz for the metric 
to simplify equations of motion. As a result, 
we end up with studying holographic model with highly symmetric restrictions on the 
field theory side.   
For example, one of the well studied system takes the metric form such as 
\bea
\label{homoisometric}
ds^2 = - a^2(r) dt^2 + \frac{dr^2}{a^2(r)} + b^2(r) (dx^2 + dy^2) \,,
\eea
which corresponds to the homogeneous and isotropic system in the dual field theory, and 
one of such solutions is, of course, well-known Schwarzschild black brane solution or 
Reissner-Nordstr\"{o}m solution in AdS. 
However in more generic setting like Einstein-Maxwell-dilaton gravity, 
we can obtain more generic 
values for $a(r)$ and $b(r)$. One such nice example is the Lifshitz geometry, 
which has a scaling symmetry but this scaling symmetry acts nontrivially. 
See, for example for the application of such generic Lifshitz-like geometries from Einstein-Maxwell-dilaton gravity to 
condensed matter system,   
\cite{Kachru:2008yh} - \cite{Gouteraux:2011qh}. 
  
However in the condensed matter systems in nature, there are systems 
which do not satisfy above homogeneity and isotropy. For example, 
due to the crystal structures of the background atomic lattices, 
some systems induce 
the anisotropic structure of the Fermi-surfaces for the electrons. 
Generically, Fermi surfaces are generically not spherically symmetric but rather quite 
anisotropic in momentum space.  
Given that such anisotropic systems are quite ubiquitous in condensed matter theory, 
obviously it is desirable to study more generic gravity solutions 
where we can relax above homogeneity and isotropy condition at IR, 
and study such systems for the holographic setting. 
In this paper we would like to study more  gravitational solutions which are homogeneous 
but not isotropic at the fixed radial slice. Putting asides the holographic applications, 
studying the generic solutions in the gravity system, 
which are homogeneous but not isotropic, is, by itself, very interesting 
problems.

Recently, by applying the Bianchi classification well-studied in cosmology, 
it is shown in \cite{Iizuka:2012iv} that 
we can obtain very generic new classes of solutions 
which are homogeneous but anisotropic in IR. 
In this paper, we restrict our 
attention to the following four-dimensional metric ansatz; 
\bea
\label{anisometric}
ds^2 =\frac{1}{z^2}\left( - g(z) dt^2 + \frac{dz^2}{g(z)} +  e^{A(z) + B(z)} dx^2  +2  c(z) dx dy +e^{A(z) - B(z)} dy^2 \right) \,,
\eea
This is the generalization of the ansatz (\ref{homoisometric})\footnote{Here we have rescaled the radial coordinate $z$ as $1/z = r$,  
such that we have explicit overall factor $1/z^2$, just for convenience.}, and admits static, 
homogeneous, but anisotropic metric  
where homogeneity at the fixed radial slice 
is achieved through two Killing vectors $\partial_x$, and $\partial_y$; 
The new function $B(z)$ and $c(z)$ are introduced to allow anisotropy. 
The point that all functions are dependent on only radial coordinate $z$, 
guarantees that this system is homogeneous at the fixed radial slice, therefore it is so 
at the dual field theory side.\footnote{We can obtain homogeneous space-time without 
two commutative Killing vectors  $\partial_x$, and $\partial_y$. In more generic situations, 
we can consider Bianchi types of the geometries where homogeneity is achieved by two non-commutative Killing vectors. See \cite{Iizuka:2012iv} for the five-dimensional analysis, where 
three spatial Killing vectors do not commute, except for the type I case. 
In this paper, we consider the simplest case, where metric admits two commutative Killing vectors  $\partial_x$, and $\partial_y$. This is analogous to the type I case studied in \cite{Iizuka:2012iv}.} 
Note that we can make the metric diagonal in $x$, $y$ place only at some fixed radius 
$z = z_i$.   
This means that by coordinate transformation for 
$x$ and $y$, we can make  $B(z_i) = c(z_i) = 0 $ 
only at some radius $z = z_i$. 
However, for 
generic metric and generic radius, we cannot make the metric diagonal in ($x$, $y$) place 
and this induces the anisotropy to the system.

The organization of this paper is followings; In section 2, we start with the analysis of
what kinds of matter profiles allow anisotropy for the homogeneous 
metric ansatz. 
Quite obviously by adding more new degrees of freedom with more generic profile, 
we can always construct more complicated anisotropic solutions. 
As many of the bottom up holographic approach does not have a definite principle for 
what degrees of freedom are needed, 
we consider the case where anisotropy is simply induced by the scalar fields. 
In section 3, we will construct both perturbatively and numerically 
some of such anisotropic solutions in the Einstein-Maxwell-dilaton system. 
In section 4, we generalize our construction to the Eistein-Maxwell-axion-dilaton system which allows $SL(2,R)$ symmetry. The good point is that once we obtain one solution, by using 
the $SL(2,R)$ symmetry, the symmetry allows us to construct another solutions. 
Section 5 is devoted to the argument for horizon area under the presence of 
anisotropic matter.  
We show how the thermodynamical third law 
holds in our set-up, implying that the horizon area goes to vanish at low temperature. 
We end with discussion at section 6.\footnote{For another approach for anisotropic black brane study, see \cite{Mateos:2011ix,Mateos:2011tv}. See also \cite{Nakamura:2009tf,Ooguri:2010kt,Ooguri:2010xs} for spontaneous homogeneity symmetry breaking in holographic approach. Recently holographic spontaneous isotropy breaking in M-theory setting is discussed in \cite{Donos:2012gg}.}

\section{Searching gravitational theories admitting anisotropy}
We start with the following question; from what kind of gravitational theories, can we have 
homogeneous but anisotropic black brane solutions on the metric form (\ref{anisometric})?
Obviously by adding more and more degrees of freedom, we can have more and more complicated 
anisotropic solutions. Therefore we first ask what is the simplest gravitational theory admitting such anisotropic solutions. 
For concreteness, we restrict out attention to the $3+1$ dimensional static homogeneous black brane solutions in asymptotically AdS spacetime.

\subsection{Generalized Gaussian Null Coordinates}

Before we analyze each system in detail, we first comment on the metric ansatz we choose. 
For the static regular homogeneous black 
branes, we can make the metric in the form written by a generalized 
gaussian null coordinates 
\bea
\label{gaussian-metric}
ds^2 &=& \frac{L^2}{z^2}\, \hat{g}_{\mu\nu}\, dx^\mu dx^\nu \nonumber \\
&=& \frac{L^2}{z^2}
\Biggl(-g(z)dv^2-2dvdz+e^{A(z)+B(z)}dx^2  +e^{A(z)-B(z)}dy^2+2c(z)dxdy \Biggr) \,,  \quad
\eea
where $B(z)$ or $c(z)$ generates the anisotropy of the metric. This metric 
can be brought back to the form (\ref{anisometric}) by coordinate transformation 
\bea
\label{nullcoord}
dv = dt - \frac{dz}{g(z)} \,,
\eea
where $g(z) > 0$ outside the horizon. 

On the generalized gaussian null coordinate, the horizon is represented as 
$z=$ constant null hypersurface where $g(z)=0$. Since the determinant of the metric 
is non-zero at the horizon, the coordinate form has advantage over the horizon such 
that there is no coordinate singularity on the horizon. In addition, 
the vector field $\p/\p z$ is the generator of a null geodesic curve and $z$ 
is the affine parameter on the conformal metric $\hat{g}_{\mu\nu}$. So, 
it is natural to assume that all the metric functions $g(z)$, $A(z)$, $B(z)$, and $c(z)$ are 
$C^2$ with respect to $z$. As shown in Appendix B, 
all the scalar curvature squares are finite if 
all the metric functions are $C^2$ and $e^{2A} - c^2 > 0$. The latter means that the physical area spanned by $x$ and $y$ are finite. Hereafter, without loss of generality, we can set 
the location of the horizon to $z=1$, then
\bea
g(z=1) = 0 \,.
\eea

\subsection{Pure Gravity}

We first consider the pure gravity system with action 
\bea
\label{action-pure-grav}
S=\int d^4x\sqrt{-g}\left( R+\frac{6}{L^2} \right) \,,
\eea
and ask if this theory admits anisotropic solution of the type we mention in the introduction. 
Here $L$ is the AdS curvature radius.

With the coordinates~(\ref{gaussian-metric}), the Einstein equations of pure gravity system 
(\ref{action-pure-grav}) can be written, after the several massage, as follows   
\bea
&& e^{2A}z^2(e^{2A}-c^2)gA''  
+e^{2A}z[g(-4e^{2A}+10c^2+zcc')+z(e^{2A}-3c^2)g']A'  
\nonumber \\
&& \quad +e^{4A}(g(6+z^2A'^2)-2zg' - 6) 
+c^2[z^2gc'^2+2zcc'(zg'-4g)-4c^2(3-3g+zg')]   \nonumber \\
&& \quad +e^{2A}[2zcc'g-z^2gc'^2-2c^2(-9+g(9+z^2A'^2)-3zg')]   =  0, 
\label{eq-A-pg} \\
&& z(c^2-e^{2A})gB''+[e^{2A}(g(2-zA')-zg') \nonumber \\ 
&& \quad -c(zgc'+c(2g(zA'-1)-zg'))]B'=0, 
\label{eq-B-pg} \\
&& z^2(e^{2A}-c^2)gc''-z[g(4c^2+e^{2A}(2+zA'))-z(e^{2A}+c^2)g']c' 
\nonumber \\
&&\,\,\,\,+[z^2gc'^2+c^2(-6+6g-2zg')  +2e^{2A}(3+(zA'-1)(3g-zg'))]c=0,  
\label{eq-C-pg} \\
&& z^2gc'^2+2zcc'(zg'-4g)-4c^2(3-3g+zg') \nonumber \\
&&\,\,\,\,+e^{2A}[12+g(-12+z(8A'-zA'^2+zB'^2)) +2z(2-zA')g']=0, 
\label{eq-Constr-pg}
\eea
where the last equation corresponds to the constraint equation.

Let us consider initial data at $z=z_i$ timelike hypersurface~($0<z_i<1$). 
By using the coordinate transformation in $(x, y)$ plane, we can make the metric at $z=z_i$ 
such that 
\bea
\label{Bzi}
B(z_i)=c(z_i) = 0 \,,
\eea
and then, $(x, y)$ plane metric becomes $ds_2^2=e^{A(z_i)}(dx^2+dy^2)$ at $z=z_i$. 
Furthermore, by an orthogonal transformation at $z=z_i$, we can rotate coordinates $(x, y)$ 
so that we can set $c'(z_i)=0$. So, initial data at $z=z_i$ timelike hypersurface is reduced to 
\bea
\label{initial-data-pg}
A(z_i), \,A'(z_i),\,B'(z_i),\,g(z_i),\,B(z_i)=c(z_i)=c'(z_i) = 0 \,. 
\eea

Since $e^{2A(z)}-c^2(z)>0$ and $g(z)>0$ for $0\le z<1$, the Eqs~(\ref{eq-A-pg})  
- (\ref{eq-C-pg}) are the regular second order differential 
equations. So, given the initial data~(\ref{initial-data-pg}), 
the solutions of the Einstein equations are uniquely determined for $0<z<1$. Now 
suppose that $c(z)=0$ for $0<z<1$. Then, substituting $c(z)=0$ into 
Eq.~(\ref{eq-B-pg}) and integrating once by $z$, one obtains 
\bea
\label{sol-B-pg}
B'(z)=\frac{C_1z^2e^{-A(z)}}{g(z)} \,, 
\eea
where $C_1$ is a constant of integration. 
By assuming regularity of the metric at the horizon $z=1$, we must set 
$C_1$ to zero, as $g(1)=0$. Therefore, with (\ref{Bzi}), 
we will obtain a solution with $B(z)=c(z)=0$, which implies isotropy. 
However due to the uniqueness of the solution, 
this is the unique solution with 
a regular event horizon of Eqs.~(\ref{eq-A-pg}) - (\ref{eq-Constr-pg}) satisfying the initial 
data~(\ref{initial-data-pg}). This means that all the solutions with a 
regular event horizon must be homogeneous and isotropic in the pure gravity 
system~\footnote{We can say that the $B(z)=c(z)=0$ solution is also unique just inside 
the event horizon. We give the argument in Appendix A.}.

To see there is a homogeneous and isotropic solution satisfying initial 
condition (\ref{initial-data-pg}) with $B(z)=c(z)=0$ in more detail,  
let us substitute $B(z)=c(z)=0$, and then 
Einstein Eqs.~(\ref{eq-A-pg}) and (\ref{eq-Constr-pg}) are 
reduced to the following two coupled differential equations 
\begin{eqnarray}
\label{eq-A_B=C=0-pg}
z^2gA''+z(zg'-4g)A'-6+g(6+z^2A'^2)-2zg' &=&0, \\
\label{eq-Constr_B=C=0-pg}
12+g(z(8A'-zA'^2)-12)+2z(2-zA')g' &= & 0, 
\end{eqnarray}
while Eqs.~(\ref{eq-B-pg}) and (\ref{eq-C-pg}) are trivially satisfied. 
Under the coordinate transformation in $(x, y)$ plane at $z=1$, we can set $A(1)=0$. 
From Eqs.~(\ref{eq-A_B=C=0-pg})$\times 2 + $(\ref{eq-Constr_B=C=0-pg}), we obtain 
\bea
2A''+A'^2=0 \,.
\eea
The solution satisfying $A(1)=0$ is 
\bea
\label{sol-A-pg}
e^A=\left(\frac{A'(1)}{2}z+\left(1- \frac{A'(1)}{2}\right)\right)^2.  
\eea
Putting back this into (\ref{eq-Constr_B=C=0-pg}) and by choosing the integration constant 
such that $g(1)=0$ is satisfied, we obtain 
\bea
\label{genericgform}
g(z) = \frac{A'(1)^2 (z-1)^3+2 A'(1) \left(z^3-3 z+2\right)+4
   z^3-4}{(A'(1)-2) (A'(1) (z-1)+2)} \,.
\eea
Now let us consider the asymptotic AdS boundary condition.  
At $z \to 0$, we have 
\bea
g(z) =  1-\frac{2 A'(1) z}{A'(1)-2}+\frac{A'(1)^2
   z^2}{(A'(1)-2)^2}+\frac{8 z^3}{(A'(1)-2)^3}+O\left(z^4\right) \,.
\eea
Therefore, putting the asymptotically AdS boundary condition at $z=0$ requires $A'(1)=0$. 
Therefore from (\ref{sol-A-pg}), we have $A(z)=0$. 
Substituting $A'(1)=0$ into Eq.~(\ref{genericgform}), 
$g(z)$ becomes 
\bea
g(z)= 1 - z^3. 
\eea
Introducing time coordinate $t$ as (\ref{nullcoord}), 
we obtain the familiar form of Schwarzschild-AdS metric, 
\bea
\label{Sch-AdS-metric}
 ds^2=\frac{L^2}{z^2}\Biggl(-(1-z^3)dt^2+\frac{dz^2}{1-z^3}+dx^2+dy^2 \Biggr). 
\eea

Thus, in this pure gravity set-up, due to the fact that the bulk degrees of freedom is tiny, it is not possible to obtain the regular solution whose horizon is regular and homogeneous but show anisotropy as ansatz (\ref{gaussian-metric}). 
Therefore we consider adding matter degrees of freedom and see if we can relax this constraint. 

\subsection{Einstein Equations with General Matter Fields}
Next we consider the system consisting of gravity and generic matter with 
the action: 
\bea
\label{action-grav-generic}
S=\int d^4x\sqrt{-g}\left(R+ \frac{6}{L^2} + {\cal L}_m(g, \p g) \right), 
\eea
where ${\cal L}_m$ are the Lagrangian density of the generic matter. The evolution 
equation for $c(z)$ in the 
metric~(\ref{anisometric}) is derived from the Einstein equations as: 
\bea
&& z^2(e^{2A}-c^2)gc''-z[g(4c^2+e^{2A}(2+zA'))-z(e^{2A}+c^2)g']c' 
\nonumber \\
&&\,\,\,\,+[z^2gc'^2+c^2(-6+6g-2zg')  +2e^{2A}(3+(zA'-1)(3g-zg'))]c 
\nonumber \\
&&\,\,\,\, \,\,\,\,+L^2(e^{2A}-c^2)(3{T^z}_z-{T^v}_v+{T^x}_x-{T^y}_y)c=-2L^2(e^{2A}-c^2){T^y}_x \,,  
\label{eq-C-general} 
\eea
where $T_{\mu\nu}\equiv-\delta {\cal L}_m(g, \p g)/\delta g^{\mu\nu}$ is the energy-momentum tensor. 
This equation tells us that $c(z)$ is always generated by the non-zero component ${T^y}_x$. 
Suppose that ${T^y}_x=0$. Then, we can obtain solutions with $c(z)=0$. 
For simplicity, hereafter, we shall restrict our search for solutions 
which satisfy $c(z) = 0$ in this 
paper.\footnote{Actually, as shown in section 2.2, we can show that $c(z)=0$ is the unique 
solution satisfying the initial data $c(z_i)=c'(z_i)=0$ in Eq.~(\ref{initial-data-pg}) when 
${T^y}_x=0$.} 

Then, substitution of $c(z)=0$ into the Einstein equations restricts ${T^\mu}_\nu$ as  
\bea
{T^\mu}_\nu=0 \qquad \mbox{for any}\quad \mu\neq \nu \quad \mbox{except}\quad {T^v}_z \,.   
\eea
The other non-trivial Einstein equations are reduced to, after several massage, 
\bea
2z^2gA''+z^2g(A'^2+B'^2)&=& 2L^2({T^t}_t-{T^z}_z) \,, \quad 
\label{eq-generic-Ein-A} \\
z^2gB''+z\{zg'+(zA'-2)g\}B' &=& L^2({T^y}_y-{T^x}_x) \,, \quad 
\label{eq-generic-Ein-B} \\
2z(zA'-2)g'+12(g-1)+zg\{A'(zA'-8)-zB'^2\} &=& 4L^2{T^z}_z \,, \quad 
\label{eq-generic-Ein-g}
\eea
where the last equation corresponds to the constraint equation. The second Eq.~(\ref{eq-generic-Ein-B}) 
implies that non-zero $B(z)$ can be generated by the anisotropic energy momentum tensor in which 
\bea
\label{Tyy-Txx}
{T^y}_y\neq {T^x}_x \,.
\eea  
In this paper, we seek the gravitational system with matter satisfying (\ref{Tyy-Txx}).

Let us consider, for example, the system consisting of a real scalar field and gravity 
with action: 
\bea
\label{action-sgrav}
S=\int d^4x\sqrt{-g}\left( R+\frac{6}{L^2}-(\nabla\phi)^2-V(\phi) \right), 
\eea
where $V(\phi)$ is the potential of the scalar field $\phi$. 
The field Eqs. of the action~(\ref{action-sgrav}) are 
\bea
\label{eq-fieldsgrav}
R_{\mu\nu}-\frac{1}{2}g_{\mu\nu}R &= & \frac{3}{L^2}g_{\mu\nu}+T_{\mu\nu} \nonumber \\
T_{\mu\nu}&=&  
\nabla_\mu\phi\nabla_\nu\phi
-\frac{1}{2}g_{\mu\nu}\left((\nabla\phi)^2+V(\phi) \right), 
\label{eq-Ein-sgrav} \\
\qquad \Box\phi &=&\frac{1}{2}V'(\phi). 
\label{eq-scalar-sgrav}
\eea

Suppose $\phi$ depends only on $z$, then 
we have ${T^x}_x={T^y}_y$ and in quite analogous to the pure gravity case, we can conclude 
that $B(z)$ must be zero. This implies that there is no 
anisotropic solution due to the uniqueness of the solution.

In the next section, we consider the matter fields such that (\ref{Tyy-Txx}) is satisfied and 
can have homogeneous but anisotropic solutions.

\section{Anisotropic black branes in Einstein-Maxwell-dilaton theory}
In this section, we analytically and numerically obtain anisotropic black brane 
solutions in Einstein-Maxwell theory coupled to a real massless scalar (dilaton)
field $\phi$. We introduce gauge potential $A_v$ in  the bulk so that it plays the role of 
chemical potential in the dual field theory. The action we consider is; 
\bea
\label{EMRS}
S=\int d^4x \sqrt{-g}\left[R+\frac{6}{L^2}
-(\nabla\phi)^2-\frac{1}{4}F_{\mu\nu}F^{\mu\nu} \right], 
\eea 
with $F_{\mu\nu}=\p_\mu A_\nu-\p_\nu A_\mu$.  
Field equations are 
\begin{eqnarray}
\label{Ein-EMRS}
 R_{\mu\nu} &=& -\frac{3}{L^2}g_{\mu\nu}+\nabla_\mu\phi\nabla_\nu\phi
+\frac{1}{2}F_{\mu\alpha}F_\nu^\alpha-\frac{1}{8}g_{\mu\nu}F^2 \,, \\ 
\label{chi-EMRS}
\Box \phi &=& 0  
 \,, \\
\label{gauge-EMRS}
 \sqrt{-g}\nabla_\nu F^{\mu\nu}&=&\p_\nu(\sqrt{-g}F^{\mu\nu})=0 \,. 
\end{eqnarray}

We shall take an ansatz for the metric~(\ref{gaussian-metric}), 
\bea
\label{gaussian-metric2}
 ds^2=\frac{L^2}{z^2}\Biggl(-g(z)dv^2-2dvdz+e^{A(z)+B(z)}dx^2  +e^{A(z)-B(z)}dy^2+2c(z)dxdy \Biggr),  
\eea
and the gauge field as 
\bea
\label{ansz-EMRS}
A_\mu dx^\mu=A_v(z)dv \,.
\eea
Just for simplicity, we furthermore restrict our attention only to the solutions which satisfy 
\bea
\label{czerocondition}
c(z) = 0 \,.
\eea

As shown in section 2, we find that the system cannot have anisotropic regular 
black brane solutions when $\phi$ has only radial coordinate dependence as 
$\phi = \phi(z)$. 
To see this, note that the energy-momentum tensor for the matter field from the action (\ref{EMRS}) is 
\bea
T^{\mu}_{\,\,\,\,\nu} = \nabla^\mu \phi \nabla_\nu \phi - \frac{1}{2} \delta^{\mu}_{\,\,\,\,\nu} (\nabla\phi)^2  + \frac{1}{2} F^\mu_{\,\,\,\, \lambda} F_{\nu}^{\,\, \lambda} -\frac{1}{8} \delta^{\mu}_{\,\,\,\,\nu}  F_{\lambda \eta}F^{\lambda \eta} \,.  
\eea
Given that gauge potential is a function of radial coordinate, we can have nonzero ${T^x}_x - {T^y}_y $ if scalar field $\phi$ has coordinate $x$- or $y$-dependence.
If we choose 
for the scalar field as \cite{Mateos:2011ix,Mateos:2011tv}, 
\bea
\label{dilatonxdependence}
\phi = \alpha x
\eea
where $\alpha$ is a constant,  then 
\bea
{T^x}_x - {T^y}_y = \alpha^2 g^{xx}  \neq 0 \,.
\eea 
Of course, the profile (\ref{dilatonxdependence}) implies $T^y_{\,\,\,\,x} = 0$. 
Therefore it is consistent with the 
assumption (\ref{czerocondition}). So,  
we consider $\alpha \neq 0$ for anisotropic solution \footnote{More explicitly, for anisotropy we need $\alpha \neq 0$ can be seen as follows; by rescaling coordinate $y$ as $y\to a y$, $B(1)$ can be set to zero. 
By Eq.~(\ref{B-EMRS}), we also obtain $B'(1)=0$. 
Then, we have initial condition  where
only one parameter is given by $A'(1)$, while $g(1) = 0$ and $g'(1)$ is determined by (\ref{g-EMRS}).  
This initial condition uniquely determines the solution, and it is straightforward to obtain  
solution with the ansatz $B(z) = 0$, which is Reissner-Nordstr\"{o}m solution in AdS. 
The regular solution obtained with $B(z) = 0$ obviously 
satisfies the boundary conditions $B(1)=B'(1)=0$, 
therefore they are the unique regular solutions of 
Eqs.~(\ref{A-EMRS}) - (\ref{g-EMRS}).}.
Note that we have introduced the manifest $x$-dependence for the scalar  
field $\phi$, but not for the metric and gauge potential. 
Therefore, metric is still homogeneous, even though  the 
scalar field $\phi$ induces inhomogeneity.

Under the ansatz, Eq.~(\ref{chi-EMRS}) is 
automatically satisfied. The solution of Eq.~(\ref{gauge-EMRS}) becomes 
\bea
\label{sol-gauge-EMRS}
F_{zv}= A_v'= L c_1 (\sqrt{-g} (g^{vz})^2)^{-1} = L c_1 e^{-A}, 
\eea
where $c_1$ is a constant corresponding to charge density of the gauge field. 
Substituting the ansatz~(\ref{ansz-EMRS}) and Eq.~(\ref{sol-gauge-EMRS}) into 
the Einstein Eqs.~(\ref{Ein-EMRS}), we obtain the following three coupled 
differential equations,
\bea
2A''+A'^2+B'^2 &=& 0, 
\label{A-EMRS} \\
zgB''+\{zg'+g(zA'-2)\}B'+\alpha^2ze^{-A-B} &= & 0,
\label{B-EMRS} \\
2z(zA'-2)g'+\{12+zA'(zA'-8)-z^2B'^2\}g && \nonumber \\
\,\,\,\, +\, c_1^2z^4 e^{-2A}+2(\alpha^2z^2e^{-A-B}-6) \, &= & 0. 
\label{g-EMRS}
\eea

Since we seek the black brane solutions with smooth horizon, 
we require all the metric functions are $C^2$.
Before we investigate these equations of motion to find explicit solutions, 
we point out that there is a difference between non-extremal case and 
extremal case of the system.  
For $\alpha \neq 0$, the above equations of motion 
at the horizon $(z=1)$, which we call ``horizon condition'' becomes 
\bea
\label{initialz1valueforAD}
2 A''(1) + A'(1)^2 + B'(1)^2 = 0 \\
\label{initialz1valueforB}
g'(1) B'(1) + \alpha^2 e^{-A(1) - B(1)} = 0 \\
\label{initialz1valueforA}
2 (A'(1) - 2) g'(1) + c_1^2 e^{-2 A(1)} + 2 (\alpha^2 e^{-A(1) - B(1)} - 6) = 0
\eea
with the assumption that $e^{A(1)}$, $e^{B(1)}$ are non-singular. 
This ``horizon condition'' 
tells the crucial difference between extremal limit 
(zero temperature limit) and 
non-extremal case (non-zero temperature). 
If black branes are extremal,  
then around the horizon $z = 1$, $g(z)$ is expected to have  
double zero, as $g \sim (1 - z)^2$.  
Then, $g' \sim (1 - z)$, and from (\ref{initialz1valueforB}), we see that  
$B'(1)$ diverges as  $1/(1-z)$, so $B(1)$ diverges as $\log (1-z)$. 
Note that these divergence contradicts with our non-singular assumption for $B(1)$,  
and this strongly supports the anisotropy 
diverges at the extremal limit. 
On the other hand, for non-extremal case, $g \sim (1-z)$, and $g' \sim$ const. 
Therefore $B'(1) $ approaches some constant, which is consistent with the non-singular assumption for $B(1)$.  
This crucial difference between non-extremal case and extremal case implies that 
extremal limit might be singular.

\subsection{Analytic solutions by perturbations}
When $\alpha$ is very small, we can obtain the analytic solutions of 
Eqs.~(\ref{A-EMRS}) - (\ref{g-EMRS}) by perturbative expansion in $\alpha$. The unperturbed 
isotropic black brane solution is the Reissner-Nordstr\"{o}m solution in AdS, 
given by 
\bea
\label{RN-AdS-metric}
ds^2 &=& \frac{L^2}{z^2}\Biggl(-g_0(z)dv^2-2dvdz+dx^2+dy^2 \Biggr), 
\nonumber \\
g_0(z)&=&1-\left(1+\frac{c_1^2}{4}\right)z^3+\frac{c_1^2}{4}z^4. 
\eea
For later convenience, let us define the non-extremal parameter $\xi$ 
as 
\bea
\label{non-extremal-pa}
c_1^2=4(\xi+\xi^2+\xi^3). 
\eea
Then, the function $g_0$ is rewritten by 
\bea
\label{sol-g0}
 g_0(z)&=&1-(1+\xi+\xi^2+\xi^3)z^3+(\xi+\xi^2+\xi^3)z^4 \nonumber \\
     &=& (1-z)(1-\xi z)\{1+(1+\xi)z+(1+\xi+\xi^2)z^2 \}, 
\eea
where the event horizon and the inner horizon are located as 
$z=1$ and $z=1/\xi$, respectively. Therefore, $0\le \xi\le 1$ and 
the extremal limit corresponds to $\xi=1$~($c_1^2=12$). 

Let us expand the anisotropic solutions of Eqs.~(\ref{A-EMRS}) - (\ref{g-EMRS}) 
around the Reissner-Nordstr\"{o}m solution in AdS as a series expansion in 
$\alpha$: 
\bea
\label{expansion}
A(z)&=&A_0(z)+\alpha^2A_1(z)+\alpha^4A_2(z)+\cdots, \nonumber \\
B(z)&=&B_0(z)+\alpha^2B_1(z)+\alpha^4B_2(z)+\cdots, \nonumber \\
g(z)&=&g_0(z)+\alpha^2g_1(z)+\alpha^4g_2(z)+\cdots,  
\eea
where $A_0(z)=B_0(z)=0$. Substituting Eqs.~(\ref{expansion}) into 
Eqs.~(\ref{A-EMRS}) - (\ref{g-EMRS}), we obtain equations of motion for 
$A_1$, $B_1$, $g_1$, as  
\bea
 A_1'' &=&0 \,, 
\label{eq-A1} \\
zg_0B_1''+(zg_0'-2g_0)B_1'+z&=&0 \,, 
\label{eq-B1} \\
g_1'-\frac{3}{z}g_1   - \frac{z}{2} &=& 0\,.
\label{eq-g1} 
\eea
The regular solutions of Eqs.~(\ref{eq-A1}), (\ref{eq-B1}), and (\ref{eq-g1}) 
are easily obtained as 
\bea
\label{expressionforA1g1}
A_1(z) = c_{A_{1a}} z + c_{A_{1b}} \quad \,, \quad g_1(z) = -\frac{z^2}{2} (1 - c_{g_1} z) \,, \\
B'_1(z) = \frac{z \left(c_{B_1} z+1\right)}{(1 - z) (1 - \xi z) \left(z
   \left(1 + \xi+ z+\xi z+ \xi^2 z\right)+1\right)} \,.
   \label{expressionforB1}
\eea
where $c_{A_{1a}}, c_{A_{1b}}, c_{B_{1}}, c_{g_{1}}$ are constants to be determined 
from the boundary condition. 

By perturbation expansion in $\alpha^2$, (\ref{initialz1valueforB}) yields 
\bea
g_1'(1) B_0'(1)  + g_0'(1) B_1'(1) + e^{-A_0(1) - B_0(1) } = 0 \,.
\eea
Using $A_0(1) = B_0(1) = B'_0(1) = 0$, and $g_0'(1) = -3 +  \xi + \xi^2 + \xi^3$, we have  
\bea
B_1'(1)  =  \frac{1}{3 -  \xi - \xi^2 - \xi^3}  \,,
\eea
for $\xi < 1$ case. 
This condition determines that $c_{B_1}$ in (\ref{expressionforB1}) must be 
\bea
c_{B_1} = -1 \,,
\eea
otherwise, $B_1(z)$ diverge as $\log (1-z)$ at the horizon $z=1$. 

In the extremal case $\xi = 1$,  we face the breakdown of perturbation as we pointed out before. 
In that case, 
\bea
g^{extremal}_{0}(z) = (1-z)^2 (1 + 2 z + 3 z^2 )  \,,
\eea
therefore, $g_0'(1)$ goes to zero as $1-z$. Then $B_1'(z)$ diverges as $1/(1-z)$, therefore,  $B_1(1)$ diverges logarithmically as $\log (1-z)$.  However this contradicts 
with the assumption that all functions are smooth at the horizon. 
Since perturbation breaks down at the extremal limit, we will consider the non-extremal 
case ($\xi < 1$) from now on. 

Since $z = 1 $ is horizon, we need $g_1(1) = 0$. This implies that in (\ref{expressionforA1g1}), 
\bea
c_{g_1} = 1 \quad \,, \quad g_1(z) = -\frac{z^2}{2} (1 -  z) \,,
\eea
Then, $g_1'(1) =  1/2$. 
Perturbation expansion in $\alpha^2$ of (\ref{initialz1valueforA}) with $A_0(z) = B_0(z) = 0$ gives
\bea
\label{initialz1forA2}
2 A_1'(1)  g_0'(1) -4 g_1'(1) 
 -2  c_1^2  A_1(1) + 2    = 0 \,.
\eea
With $g_1'(1) =  1/2$, (\ref{initialz1forA2}) becomes, 
\bea
\label{CA1aCA1brelation}
 (-3+  \xi + \xi^2 + \xi^3 ) c_{A_{1a}} = 4 (\xi + \xi^2 + \xi^3)  c_{A_{1b}}  \,.
\eea
Finally we impose boundary condition at $z=0$ that the solution is asymptotic to 
AdS. 
We can require the boundary condition such that 
\bea
A_1(0) = A_1'(0) = 0 \,,
\eea
which is consistent with (\ref{CA1aCA1brelation}). 
This ends up determining all the constant 
and we finally obtain, 
\bea
\label{sol1}
 A_1(z)&=&0, \qquad g_1(z)=-\frac{z^2}{2}(1-z), \nonumber \\
 B_1(z)&=&
 -\frac{2\ln(1-\xi z)-\ln\{1+(1+\xi)z+(1+\xi+\xi^2)z^2\}}{2(1+2\xi+3\xi^2)} \nonumber \\
&\quad &+ \frac{1 + 3 \xi}
{(1+2\xi+3\xi^2)\sqrt{3+2\xi+3\xi^2}} \times \nonumber \\
&& \left(
{  
\arctan\left(\frac{1+\xi}{\sqrt{3+2\xi+3\xi^2}}\right)
-\arctan\left(\frac{1+\xi+2(1+\xi+\xi^2)z}{\sqrt{3+2\xi+3\xi^2}}\right)   
}
 \right) \,. \quad \quad 
\eea
This satisfies asymptotic AdS condition  
$B_1(0) = B_1'(0) = 0$ 
as 
\bea
 B_1(z) = \frac{z^2}{2 } + O(z^3)  \,.  \quad (z \to 0) 
\eea

Similarly, we can easily show that the second order solutions, $A_2$, $B_2$, and $g_2$ are 
also regular outside and on the horizon as follows. 
Substituting Eqs.~(\ref{expansion}) into 
Eqs.~(\ref{A-EMRS}) - (\ref{g-EMRS}), we obtain equations of motion for 
$A_2$, $g_2$, and $B_2$ as 
\bea
 A_2'' &=& -\frac{B_1'^2}{2} \,, 
\label{eq-A2} \\
 g_2'-\frac{3}{z}g_2 &= & \frac{zA_2'g_0'}{2}-\left(2A_2'+\frac{zB_1'^2}{4}\right)g_0
-\frac{zB_1}{2}  
\equiv S_g \,,  
\label{eq-g2} \\
 B_2''+\left(\frac{g_0'}{g_0}-\frac{2}{z}\right)B_2' &=&\frac{B_1}{g_0}
+\left(\frac{2g_1}{zg_0}-\frac{g_1'}{g_0}\right)B_1'-\frac{g_1}{g_0}B_1''  
\equiv S_B \,. 
\label{eq-B2}
\eea

By imposing $A_2(0)=A_2'(0)=0$ at the boundary 
condition for asymptotically AdS spacetime, we can formally obtain the solution $A_2$ by 
Eq.~(\ref{eq-A2}) as 
\bea
\label{sol-A2}
A_2(z)=-\frac{1}{2}\int_0^z\left(\int^{z'}_0 B_1'(z'')^2dz'' \right)dz', 
\eea
which is obviously regular outside and on the horizon, $z=1$, as $B_1'$ is finite there. This 
solution means that the horizon area per unit coordinate interval, $\Delta x=\Delta y=1$, 
$e^{A(1)}$ decreases due to the existence of anisotropy, $B_1'\neq 0$. 

The solution of Eq.~(\ref{eq-g2}) is also obtained by the boundary condition $g_2(1)=0$ 
as 
\bea 
g_2(z)=z^3\int^z_1\frac{S_g(z')}{z'^3}dz'. 
\eea
Since $S_g$ is regular outside and on the horizon, and decays as $S_g\sim z^3$ near the boundary 
of asymptotically AdS spacetime, $g_2$ is also regular there. 
Finally, $B_2$ is also obtained by integrating 
Eq.~(\ref{eq-B2}) as 
\bea
B_2'(z)=\frac{z^2}{g_0(z)}\int^z_1\frac{g_0(z')}{z'^2}S_B(z')dz', 
\eea 
where we imposed a regular boundary condition on the horizon. Using the fact 
that $S_B$ is regular outside and on the horizon and it decays as $S_B\sim z^2$ 
near the AdS boundary, we find that $B_2'$ is also regular there and $B_2'(0)=0$. 
$B_2$ is obtained by imposing the boundary condition $B_2(0)=0$ as 
\bea
B_2(z)=\int^z_0\left(
\frac{z'^2}{g_0(z')}\int^{z'}_1\left(\frac{g_0(z'')}{z''^2}S_B(z'')dz''\right)dz' 
\right). 
\eea
Thus, we have checked that the perturbed solutions in Eqs.~(\ref{expansion}) expanded 
in a series of $\alpha^2$ satisfy the boundary condition near the AdS boundary: 
\bea
A(z)=O(z^2), \qquad B(z)=O(z^2), \qquad g(z)=1+O(z^2), 
\eea
and they are regular outside and on the horizon, 
up to the second order $O(\alpha^4)$.

We can in principle continue working on higher $\alpha^2$ corrections and obtain anisotropic solutions 
by $\alpha^2$ perturbation in this way.

\subsection{Numerical solutions}
We can also numerically obtain anisotropic black brane solutions by solving 
Eqs.~(\ref{A-EMRS}) - (\ref{g-EMRS}).  
By rescaling $x$ and $y$ properly, $A(1)$ and $B(1)$ can be set to zero. 
Under the conditions, we can read off the regularity condition at the 
horizon from ``horizon condition'' Eqs.~(\ref{initialz1valueforAD}) - (\ref{initialz1valueforA}) as 
\bea
\label{initcon-EMRS}
A'(1)=2+\frac{\alpha^2-6}{\kappa}+\frac{c_1^2}{2\kappa} \,, \qquad 
B'(1)=\frac{\alpha^2}{\kappa} \,, 
\eea   
where $\kappa$ is defined as $\kappa\equiv-g'(1)$. 
Thus, the regular solutions of Eqs.~(\ref{A-EMRS}) - (\ref{g-EMRS}) 
are uniquely determined by three parameters, $\alpha$, $c_1$, and $\kappa$.  

As the boundary conditions for the asymptotically AdS spacetime, we require 
that 
\bea 
\label{bd-asym-EMRS}
A(0)=B(0)=A'(0)=B'(0)=0 \,. 
\eea
Under the boundary conditions, one obtains the asymptotic behavior of $g(z)$ by 
Eq.~(\ref{g-EMRS}) as 
\bea
g(z)=1+O(z^2) \,.  
\eea

The first two conditions of Eqs.~(\ref{bd-asym-EMRS}) are automatically 
satisfied by rescaling $x$ and $y$ again for the solutions of 
Eqs.~(\ref{A-EMRS}) - (\ref{g-EMRS}) satisfying the regularity conditions~(\ref{initcon-EMRS}), 
\bea
\label{rescaling}
x\to e^{-(A(0)+B(0))/2}x \,, \quad y\to e^{-(A(0)-B(0))/2}y \,, \quad  
\alpha \to \alpha e^{-(A(0)+B(0))/2} \,. 
\eea
So, the only task is to obtain the solutions satisfying the latter two conditions, 
$A'(0)=B'(0)=0$. $B'(0)=0$ is automatically satisfied because 
the asymptotic behavior of $B$ is derived from Eq.~(\ref{B-EMRS}) as
\bea
\label{asy-B-EMRS}
B(z)=B(0)+O(z^2) \,. 
\eea
Thus, the anisotropic black brane solutions in asymptotically AdS spacetime are obtained 
by searching the parameters $\alpha$, $c_1$, and $\kappa$ satisfying $A'(1)=0$. 
 
We numerically find the value $c_1$ satisfying $A'(0)=0$ for a fixed $\alpha$ 
and $\kappa$ by solving Eqs.~(\ref{A-EMRS}) - (\ref{g-EMRS}) from the horizon~($z=1$) 
to the infinity~($z=0$) under the regularity condition~(\ref{initcon-EMRS}). 
Thus, the numerical solutions rescaled by Eq.~(\ref{rescaling}) always 
satisfy the boundary conditions~(\ref{bd-asym-EMRS}). 

\begin{figure}
 \begin{center}
  \includegraphics[width=7truecm,clip]{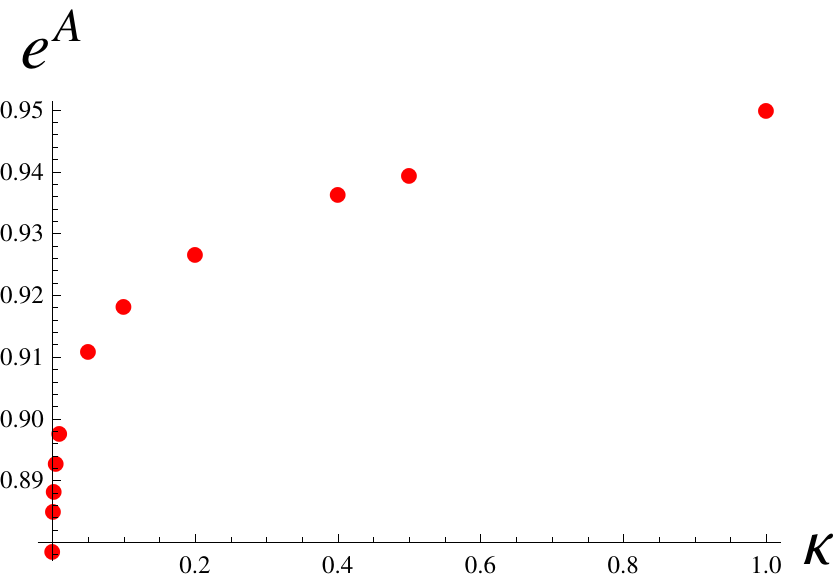}
  \caption{(color online) $\kappa$-$e^{A(1)}$ relation. The area of the horizon per unit 
$\Delta x=\Delta y=1$ are shown as a function $\kappa$ for $\alpha=\sqrt{2}$. } 
 \end{center}
\end{figure}
\begin{figure}
 \begin{center}
  \includegraphics[width=7truecm,clip]{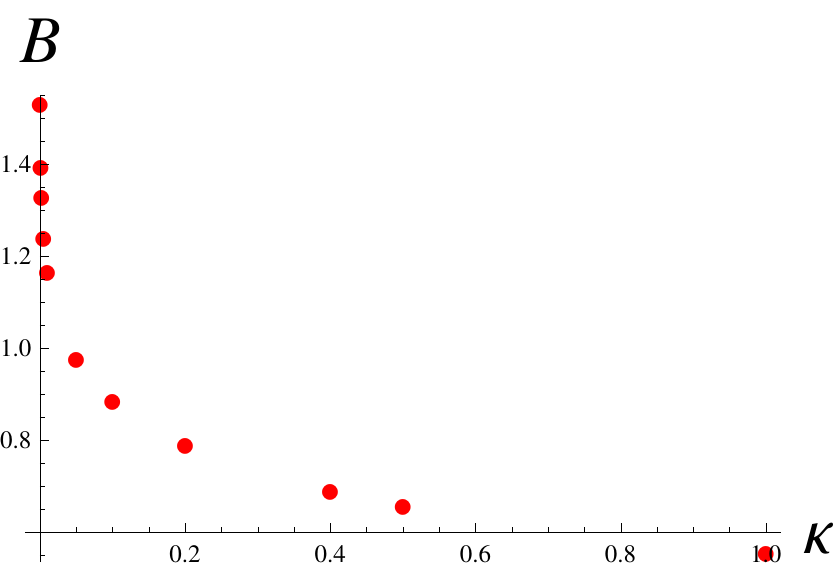}
  \caption{(color online) $\kappa$-$B(1)$ relation for $\alpha=\sqrt{2}$. 
$B(1)$ rapidly grows as $\kappa$ decreases to zero. } 
 \end{center}
\end{figure}
\begin{figure}
 \begin{center}
  \includegraphics[width=7truecm,clip]{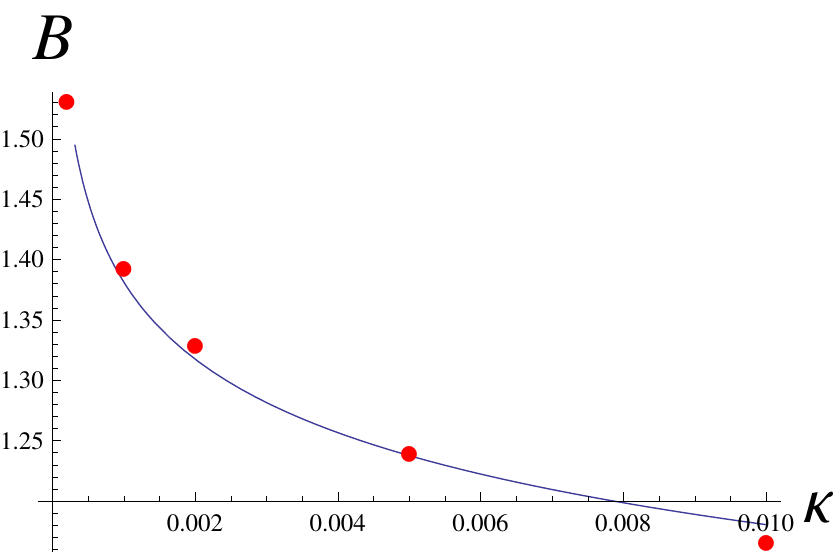}
  \caption{(color online) $\kappa$-$B(1)$ relation near $\kappa=0$ 
for $\alpha=\sqrt{2}$. The solid curve $B(1)=0.862\,\kappa^{-0.068}$ fits 
the plot well. } 
 \end{center}
\end{figure}

Figs.~1 - 6 show the anisotropic black brane solutions for the rescaled coordinates. 
Figs.~1 - 2 plot the horizon area $e^{A(1)}$ and $B(1)$ as a function of 
$\kappa$~(the minimum of $\kappa$ is $2\times 10^{-4}$) for 
$\alpha=\sqrt{2}$. As shown in Fig.~3, the curve $B(1)=0.862\,\kappa^{-0.068}$ fits 
the plot of Fig.~3 well. This suggests that $B(1)$ diverges as $\kappa$ goes to zero. 
According to the divergence, anisotropy grows near the horizon as the solution 
approaches extremal~($\kappa\to 0$). Figs.~4 - 6 show the metric as a function 
of $z$ for each $\kappa$ and $\alpha=\sqrt{2}$.  

\begin{figure}
 \begin{center}
  \includegraphics[width=6.6truecm,clip]{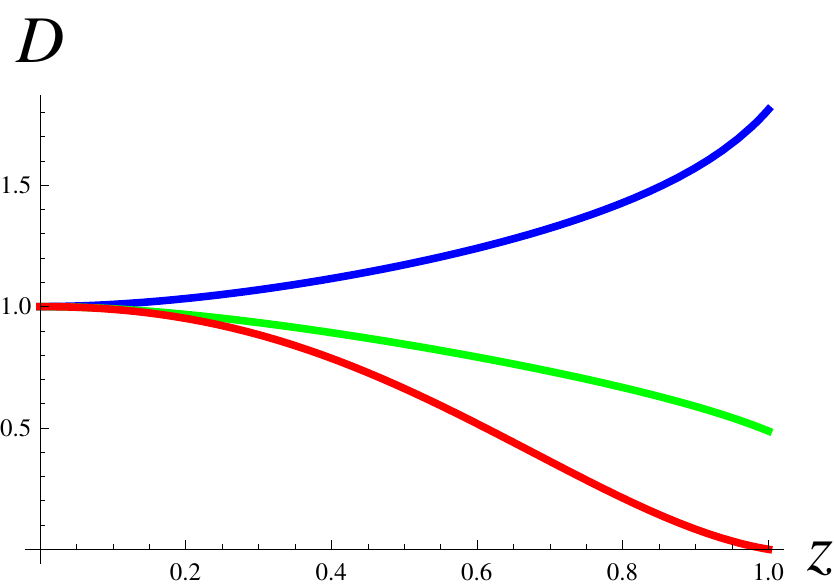}
  \caption{(color online) $D=z^2g_{xx}/L^2$~(blue), $z^2g_{yy}/L^2$~(green), and 
  $g$~(red) are shown as a function of $z$ for $\alpha=\sqrt{2}$, $\kappa=0.5$. } 
 \end{center}
\end{figure}
\begin{figure}
 \begin{center}
  \includegraphics[width=6.6truecm,clip]{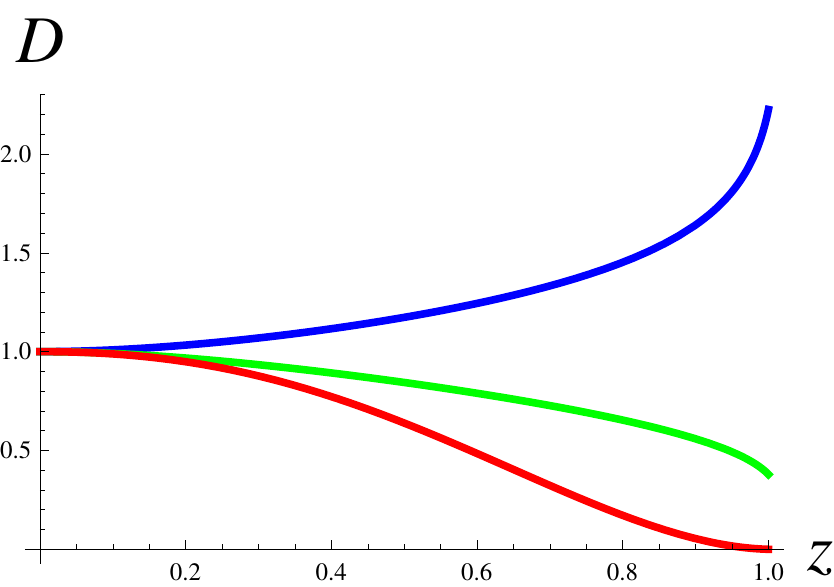}
  \caption{(color online) $D=z^2g_{xx}/L^2$~(blue), $z^2g_{yy}/L^2$~(green), and 
  $g$~(red) are shown as a function of $z$ for $\alpha=\sqrt{2}$, $\kappa=0.1$. } 
 \end{center}
\end{figure}
\begin{figure}
 \begin{center}
  \includegraphics[width=6.6truecm,clip]{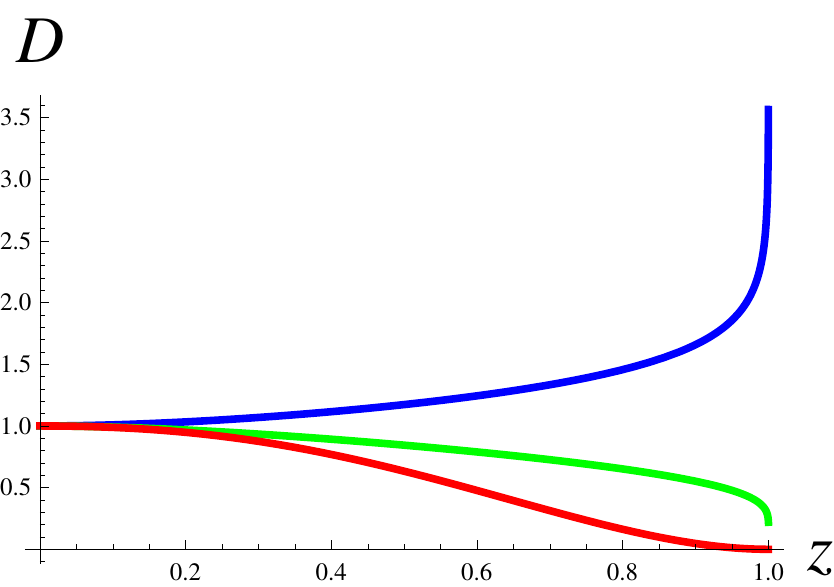}
  \caption{(color online) $D=z^2g_{xx}/L^2$~(blue), $z^2g_{yy}/L^2$~(green), and 
  $g$~(red) are shown as a function of $z$ for $\alpha=\sqrt{2}$, $\kappa=0.001$. } 
 \end{center}
\end{figure}

By Eqs.~(\ref{A-EMRS}) - (\ref{g-EMRS}), we can argue that the horizon area 
$e^{A(1)}$ per unit length, $\Delta x=\Delta y=1$ 
must go to zero as $\kappa\to 0$. Eq.~(\ref{A-EMRS}) is rewritten in the form of 
Raychaudhuri equation for the null geodesics:
\bea
\label{Raychaudhuri-EMRS}
\theta'=-\frac{1}{2}\theta^2-\frac{1}{2}B'^2 \,, 
\eea
where $\theta$ is the expansion of the congruence defined as $\theta \equiv A'$. Integrating this 
by $z$ twice, we obtain the inequality
\bea
\label{inequality-EMRS}
 A(z)\le -\int^{z}_0\int^{z'}_0 \frac{1}{2} \, B'(z'')^2 dz''dz' \,,  
\eea
where we used $A(0)=0$ to remove the integration constant. 
As $B(1)$ diverges at the extremal limit, $B(z)$ would diverge logarithmically as  
$B\sim -\ln(1-z)$. This implies that the r.h.s. of Eq.~(\ref{inequality-EMRS}) 
also negatively diverges at least logarithmically, and then $A(1)\to -\infty$, or $e^{A(1)}\to 0$.

\section{Einstein-Maxwell-dilaton-axion SL(2,R) model}
In this section, we consider Einstein-Maxwell theory coupled to a dilaton-axion 
with the action 
\bea
\label{action-SL2}
S=\int d^4x\sqrt{-g}\Biggl(R+\frac{6}{L^2}-2(\nabla\phi)^2
-\frac{1}{2}e^{4\phi}(\nabla a)^2  -e^{-2\phi}F^2-aF\tilde{F} \Biggr) \,. 
\eea 
It is well known that this action is invariant under global $SL(2,R)$ transformations \cite{Shapere:1991ta},
\bea
\label{SL(2,R)-transformation1}
\lambda\to \frac{\tilde{a}\lambda+b}{c\lambda+d} \,, \qquad \tilde{a}d-bc=1 \,, 
\eea
where $\lambda \equiv a+ie^{-2\phi} \equiv \lambda_1 + i \lambda_2$ and 
\bea
\label{SL(2,R)-transformation2}
F_{\mu\nu} \to \left(c \lambda_1 + d \right)F_{\mu\nu} - c \lambda_2 \tilde F_{\mu\nu} \,.
\eea
Therefore, once we obtain a charged black brane 
solution of the theory, we also obtain both electrically and magnetically charged 
black brane solutions by the $SL(2,R)$ transformations. 
This model is studied in holographic setting 
for the Quantum Hall effects in \cite{Goldstein:2010aw,Bayntun:2010nx}. 
At the quantum level, this $SL(2,R)$ symmetry will be enhanced to $SL(2,Z)$. For the connection of $SL(2,Z)$ to Quantum Hall Effects, 
see \cite{Witten:2003ya, Shapere:1988zv, Lutken:1991jk, Kivelson:1992zz, Burgess:2000kj} 
and review \cite{dolan:2006zc}.

In the purely electrically charged case, the last term in Eq.~(\ref{action-SL2}) 
vanishes and  then, the equations of motion are 
\bea
&& R_{\mu\nu} = -\frac{3}{L^2}g_{\mu\nu}+2\nabla_\mu\phi\nabla_\nu\phi
+\frac{1}{2}e^{4\phi}\nabla_\mu a\nabla_\nu a    
+ 2e^{-2\phi}F_{\mu\lambda}{F_\nu}^\lambda-\frac{1}{2}g_{\mu\nu}e^{-2\phi}F^2 \,, \quad
\label{eq-Ein-SL2} \\
&& \Box\phi-\frac{1}{2}e^{4\phi}(\nabla a)^2+\frac{1}{2}e^{-2\phi}F^2 = 0 \,, 
\label{eq-dilaton-SL2} \\
&& \Box a+4\nabla_\mu\phi\nabla^\mu a=0 \,, 
\label{eq-axion-SL2} \\
&& \sqrt{-g}\nabla_\mu(e^{-2\phi}F^{\mu\nu}) = \p_\mu(\sqrt{-g}e^{-2\phi}F^{\mu\nu}) =0 \,. 
\label{eq-F-SL2}
\eea
We take an ansatz 
\bea
\label{SL2Zansatzfora}
\phi=\phi(z), \qquad a=\alpha x, \qquad A_\mu dx^\mu=A_vdv, \qquad c(z) = 0 \,,
\eea
under the metric~(\ref{gaussian-metric2}).
Eq.~(\ref{eq-axion-SL2}) is automatically satisfied and 
the solution of Eq.~(\ref{eq-F-SL2}) is written by 
\bea
\label{sol-gauge-SL2}  
F_{zv} = A_v'= -c_2Le^{-A}e^{2\phi} \,, 
\eea
where $c_2$ is a constant corresponding to 
the charge density of the gauge field. 
The energy-momentum  tensor is 
\bea
T^{\mu}_{\,\,\,\,\nu} &=& 2 \nabla^\mu \phi \nabla_\nu \phi 
+\frac{1}{2}e^{4\phi}\nabla^\mu a\nabla_\nu a    
+ 2e^{-2\phi}F_{\mu\lambda}{F_\nu}^\lambda \nonumber \\
&& -\frac{1}{2} \delta^\mu_{\,\,\,\,\nu} \left( 
2(\nabla\phi)^2
+ \frac{1}{2}e^{4\phi}(\nabla a)^2
+ e^{-2\phi}F^2  \right) \,.
\eea
Then, ansatz (\ref{SL2Zansatzfora}) satisfies $T^x_{\,\,\,\,x} \neq T^y_{\,\,\,\,\,y}$ and $T^y_{\,\,\,\,x} = 0$, therefore we expect to have a $B(z) \neq 0$ and $c(z) = 0$ solution. 

Substituting Eq.~(\ref{sol-gauge-SL2}) into 
Eqs.~(\ref{eq-Ein-SL2}) and (\ref{eq-dilaton-SL2}), we obtain four-coupled differential equations, 
\bea
 2A''+A'^2+B'^2+4\phi'^2 &=& 0 \,, 
\label{eq-A-SL2} \\
 zgB''+\{zg'+g(zA'-2)\}B'+\frac{1}{2}\alpha^2e^{4\phi-A-B}z&=&0 \,, 
\label{eq-B-SL2} \\
 2z(2-zA')g'-\{12-8zA'+z^2(A'^2-B'^2-4\phi'^2)\}g \quad &&\nonumber \\
+12-\alpha^2z^2e^{4\phi-A-B}-4c_2^2z^4e^{2\phi}e^{-2A}&=&0 \,, 
\label{eq-g-SL2} \\
 e^Ag\phi''+z^2\left(\frac{e^Ag}{z^2} \right)'\phi'
-\frac{1}{2}\alpha^2e^{4\phi}e^{-B}-c_2^2z^2e^{-A}e^{2\phi}&=&0 \,, 
\label{eq-phi-SL2}
\eea
where the third equation corresponds to the constraint equation. 
The equations of motion (\ref{eq-A-SL2}) - (\ref{eq-phi-SL2}) are 
invariant under the transformation
\bea
\label{trans-dilaton-SL2}
\phi\to \phi-\phi_0 \,, \quad  
\alpha\to e^{2\phi_0}\alpha \,, \quad  
c_2\to e^{\phi_0}c_2 \,,
\eea
for an arbitrary value of $\phi_0$. This is nothing but the $SL(2,R)$ symmetry 
given by (\ref{SL(2,R)-transformation1}) and (\ref{SL(2,R)-transformation2}) with $SL(2,R)$ parameter 
$\tilde{a} = d^{-1} = e^{\phi_0}$, $ b = c= 0$.
Using the freedom of this transformation~(\ref{trans-dilaton-SL2}) and 
the rescaling of the coordinate $x$ and $y$, we can set 
\bea
\label{initial-cond-SL2}
A(1)=B(1)=\phi(1)=g(1)=0 \,. 
\eea
By Eqs.~(\ref{eq-A-SL2}) - (\ref{eq-phi-SL2}), the regularity of the black brane solution 
requires 
\bea
\label{regularity-SL2}
 A'(1)=2+\frac{\alpha^2-12+4c_2^2}{2\kappa} \,, \quad  
 B'(1)=\frac{\alpha^2}{2\kappa} \,, \quad 
\phi'(1)=-\frac{2c_2^2+\alpha^2}{2\kappa},  
\eea
where $g'(1)=-\kappa$. Thus, the regular solutions of 
Eqs. (\ref{eq-A-SL2}) - (\ref{eq-phi-SL2}) are uniquely determined by 
three parameters, $\alpha$, $c_2$, and $\kappa$. We consider 
the boundary condition (\ref{bd-asym-EMRS}) adopted in section 3. 
For simplicity, we also require that $\phi=0$ at the 
infinity. Since the asymptotic behavior of $B$ is given by 
Eq. (\ref{asy-B-EMRS}), we can set $A(0)=B(0)=B'(0)=0$ by the 
coordinate transformation~(\ref{rescaling}). So, we can numerically 
find the value of $c_2$ satisfying $A'(0)=0$ at the infinity for a 
given $\kappa$ and $\alpha$. Once the solutions are obtained, 
we can set $\phi(0)=0$ by the transformation (\ref{trans-dilaton-SL2}).  

\begin{figure}
 \begin{center}
  \includegraphics[width=7truecm,clip]{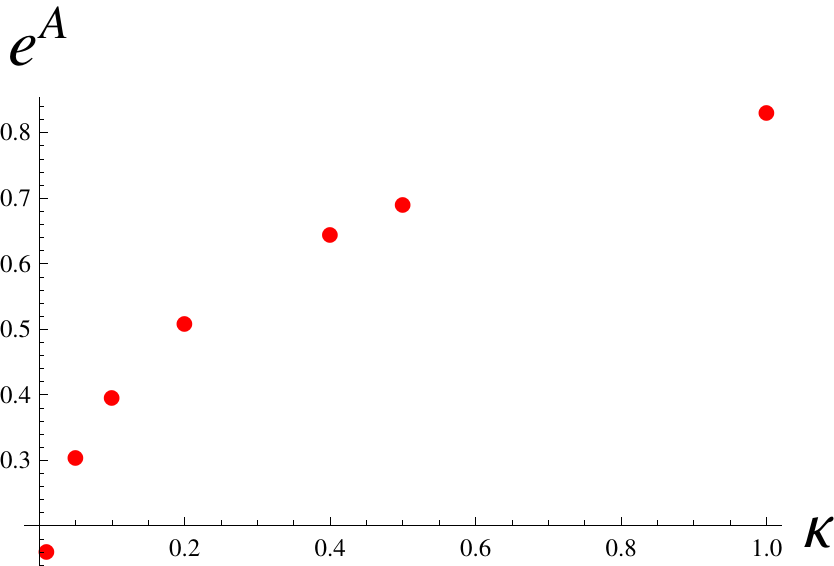}
  \caption{(color online) $\kappa$-$e^{A(1)}$ relation. The area of the horizon per unit 
$\Delta x=\Delta y=1$ are shown as a function $\kappa$ for $\alpha=\sqrt{2}$. The 
area rapidly decreases to zero as $\kappa\to 0$.} 
 \end{center}
\end{figure}
\begin{figure}
 \begin{center}
  \includegraphics[width=7truecm,clip]{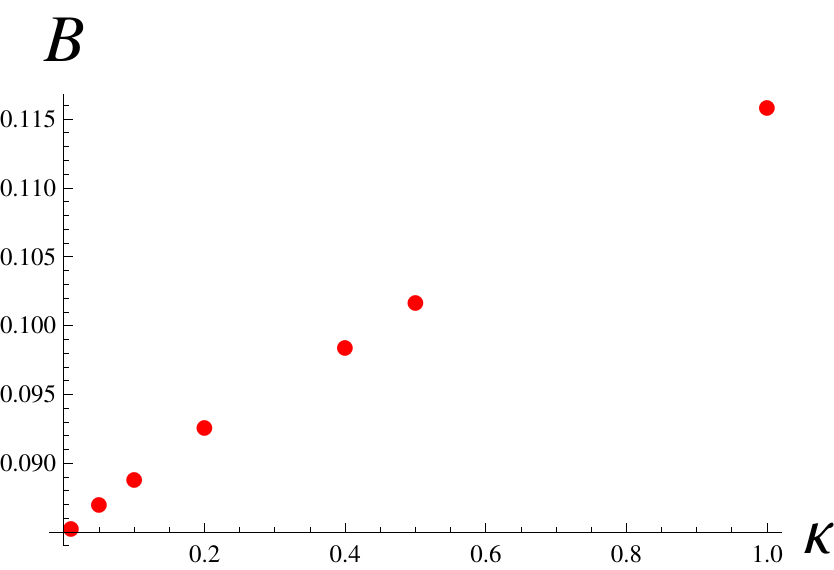}
  \caption{(color online) $\kappa$-$B(1)$ relation for $\alpha=\sqrt{2}$. 
$B(1)$ goes to zero as $\kappa$ decreases to zero. } 
 \end{center}
\end{figure}
\begin{figure}
 \begin{center}
  \includegraphics[width=7truecm,clip]{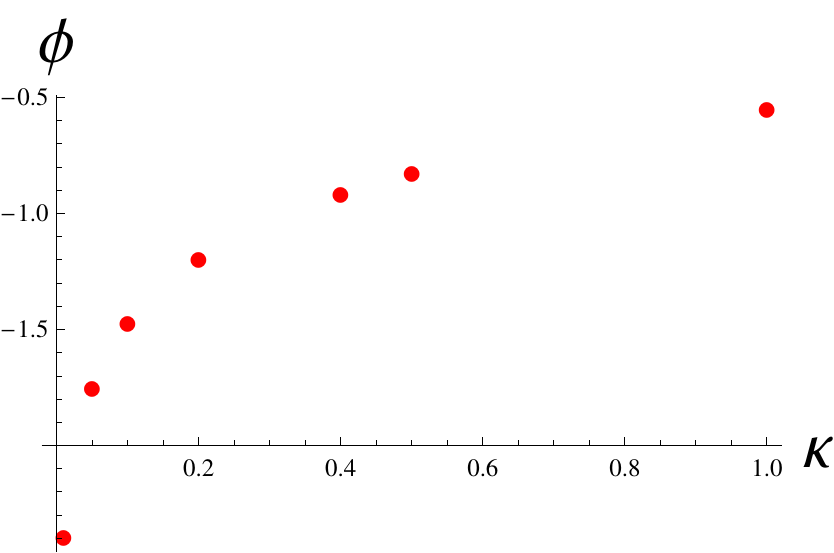}
  \caption{(color online) $\kappa$-$\phi(1)$ relation for 
$\alpha=\sqrt{2}$. $\phi(1)$ negatively diverges as $\kappa$ 
decreases to zero.} 
 \end{center}
\end{figure}
Figs.~7 - 12 show the anisotropic black brane solutions for the rescaled coordinates. 
Figs.~7 - 9 plot the horizon area $e^{A(1)}$, and the values of $B$, $\phi$ at 
the horizon as a function of $\kappa$~(the minimum of $\kappa$ is $0.01$) 
for $\alpha=\sqrt{2}$. Due to the existence of the dilaton field $\phi$, the 
horizon area rapidly decreases to zero at the extremal limit, 
$\kappa\to 0$~(Fig.~7), and the anisotropy also decreases~(Fig.~8.), contrary 
to the previous real scalar model. This is because the negative divergence of the 
dilaton field strongly suppresses the kinetic term 
$e^{4\phi}(\nabla a)^2$ in Eq.~(\ref{action-SL2}) which 
generates the anisotropy.  

\begin{figure}
 \begin{center}
  \includegraphics[width=6.6truecm,clip]{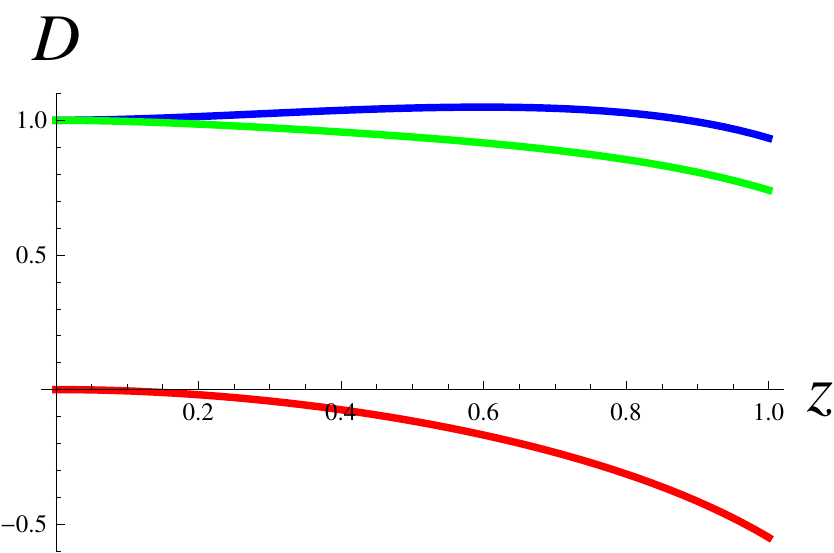}
  \caption{(color online) $D=z^2g_{xx}/L^2$~(blue), $z^2g_{yy}/L^2$~(green), and 
  $\phi$~(red) are shown as a function of $z$ for $\alpha=\sqrt{2}$, $\kappa=1$. } 
 \end{center}
\end{figure}
\begin{figure}
 \begin{center}
  \includegraphics[width=6.6truecm,clip]{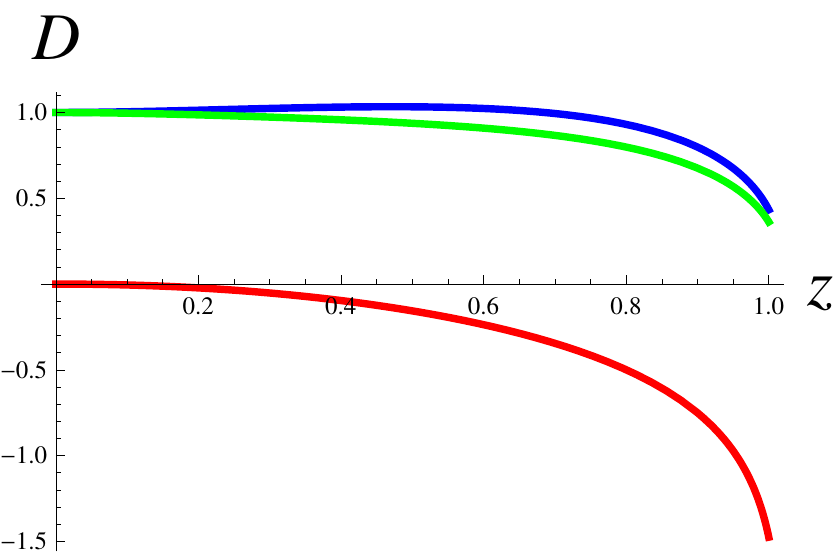}
  \caption{(color online) $D=z^2g_{xx}/L^2$~(blue), $z^2g_{yy}/L^2$~(green), and 
  $\phi$~(red) are shown as a function of $z$ for $\alpha=\sqrt{2}$, $\kappa=0.1$. } 
 \end{center}
\end{figure}
\begin{figure}
 \begin{center}
  \includegraphics[width=6.6truecm,clip]{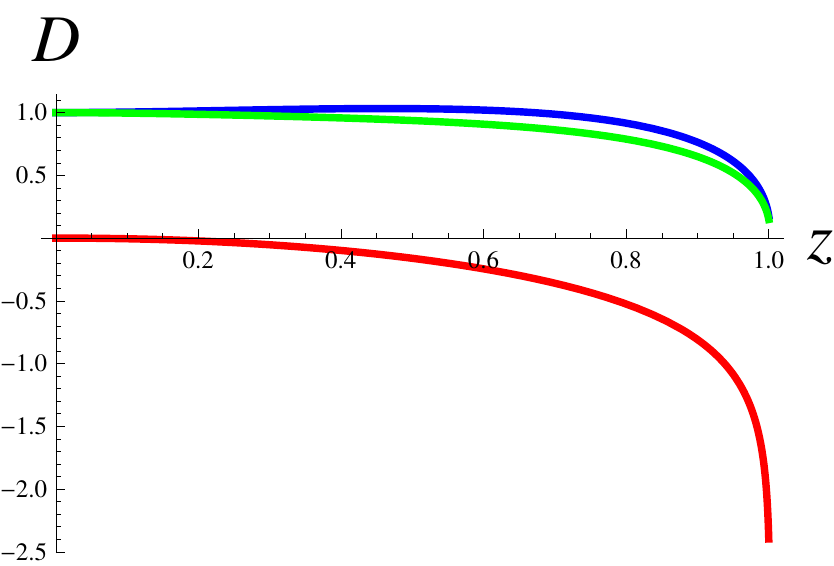}
  \caption{(color online) $D=z^2g_{xx}/L^2$~(blue), $z^2g_{yy}/L^2$~(green), and 
  $\phi$~(red) are shown as a function of $z$ for $\alpha=\sqrt{2}$, $\kappa=0.01$. } 
 \end{center}
\end{figure}
Figs.~10 - 12 show the metric as a function 
of $z$ for $\kappa=1, 0.1, 0.01$ and $\alpha=\sqrt{2}$.  
We can see that the difference between $g_{xx}$ and $g_{yy}$ 
goes to zero as $\kappa$ decreases to zero. This means that 
the geometry near the horizon becomes isotropic when the black 
hole approaches the extremal solution. 

Given the solution shown in Figs.~7 - 12, then by $SL(2,R)$ symmetry, 
we can obtain more generic solutions where all the fields, 
{\it i.e.,} dilaton $\phi$, axion $a$, 
field strength $F_{zv}$ and $F_{xy}$ have more generic position $x$-dependence. 
However, since $SL(2,R)$ symmetry does not transform the metric, the 
metric is still anisotropic but homogenous. This means that the position dependence 
of $\phi$, $a$, $F_{zv}$, $F_{xy}$ does not yield the position dependence for the 
energy-momentum tensor.

\section{The third law of anisotropic black branes}

In this section, we consider the third law of thermodynamics for the general 
ainsotropic black brane solutions. The third law states that the area of the 
black brane horizon goes to zero as the temperature decreases to zero. As 
seen in Eq.~(\ref{inequality-EMRS}), the area must go to zero when $B(z)$ 
diverges. Indeed, we can give a theorem supporting the third law of thermodynamics.

We consider the following static anisotropic black brane solutions with the 
metric 
\bea
\label{Gaussian-nullMetric1}
ds^2=\frac{L^2}{z^2}\Bigl(-g(z)dv^2-2dvdz+e^{A(z)+B(z)}dx^2+ e^{A(z)-B(z)}dy^2 \Bigr). 
\eea 
As the asymptotic boundary condition, we shall impose that 
\bea
\label{boundary-con}
\lim_{z\to 0}A(z)=\lim_{z\to 0}B(z)=\lim_{z\to 0}A'(z)=\lim_{z\to 0}B'(z)=0.    
\eea

In the neighbourhood of the horizon, we assume that 
there is a small positive value $\epsilon$ such that in $1-\epsilon <z\le 1$, 
$g$ can be expanded as
\bea
\label{near-horizon-g}
g(z)=\kappa (1-z)+g_2(\kappa)(1-z)^2+O(\epsilon^3). 
\eea
Note that on top of the horizon where $z\to 1$, the first term always dominates 
over the second term. However, slightly away from the horizon, like $z = 1 - \epsilon$, with 
$\epsilon > 0$, then, near the extremal limit where $\kappa \to 0$, the second term 
dominates over the first term.  
Given this, then we can show the following theorem: \\
\\
{\it Theorem}. \\
For the anisotropic black brane solutions with metric~(\ref{Gaussian-nullMetric1})  
satisfying the Einstein equations, if the following two conditions, 
\begin{enumerate}
\item The null convergence condition is satisfied, {\it i.e.}, $R_{\mu\nu}V^\mu V^\nu\ge 0$ 
for any null vector $V^\mu$, 
\item 
Defining 
\bea
\Delta T(z) \equiv \frac{L^2}{z^2} \, (T^y_{\,\,\,\,y}(z)-T^x_{\,\,\,\,x}(z)) = e^{-A + B }T_{yy} - e^{-A-B} T_{xx}
\eea 
as the difference for the 
re-normalized\footnote{$L^2/z^2$ is due to the overall metric factor} energy-momentum-tensor, 
it is expanded as 
\bea
\label{con-deviation-energy}
\Delta T(z)=\alpha(\kappa)+\gamma(\kappa)(1-z)+O(\epsilon^2)
\eea
in $1-\epsilon <z\le 1$ and 
$\lim_{\kappa\to +0}\alpha(\kappa)=\alpha_0~(\neq 0)$, 
$\lim_{\kappa\to +0}\gamma(\kappa)=\gamma_0$, 
\end{enumerate}
are satisfied, the horizon area must go to zero as $\kappa$ decreases to zero. 

The second condition implies that anisotropic matters, which induces nonzero $B(z)$, exist near the horizon. 
We shall prove the theorem by using the following Lemma. \\
\\
{\it Lemma}.  \\
$A$ is non-positive and $A$ and $u \equiv e^A$ satisfy the following inequalities, 
\bea
\label{con-u}
0\le u(z)\le 1, \qquad \theta=\frac{u'}{u}=A'\le 0.   
\eea  
{\it proof}) \\
From the Einstein equations, we obtain 
\bea
\label{eq-Raychau1}
\theta'=-\frac{1}{2}\theta^2-T_{zz}-\frac{1}{2}B'^2.  
\eea
From the condition~1, we have $T_{zz}=R_{zz}\ge 0$. Thus, $A'$ must be a 
non-increasing function. By the asymptotic boundary 
condition~(\ref{boundary-con}), {\it i.e.}, $A(0)=A'(0)=0$, $A'\le 0$ and $A\le 0$.  
In addition, by definition of $u$, we have $0\le u(z)\le 1$. $\Box$. \\
\\
{\it proof of Theorem}). \\
Let us suppose that the area at the horizon does not go to zero at the extremal limit, 
\bea
\label{hypo1}
\lim_{\kappa\to +0}A(1;\kappa)=A_0. 
\eea
Integrating (\ref{eq-generic-Ein-B}) by $z$, we obtain 
\bea
\label{sol-B}
B'(z)=-\frac{z^2e^{-A(z)}}{g(z)}
\int^1_z\frac{\Delta T(z')}{z'^2}e^{A(z')}dz', 
\eea 
where the constant of integration is determined by imposing regularity at the horizon.  

When $\kappa$ is small enough, by the condition 2, the sign of $\Delta T(z)$ does not 
change for $1-\epsilon <z\le 1$. So, $|B'(z)|$ can be evaluated for 
$1-\epsilon <z\le 1$ as 
\bea
\label{inequality-B'}
 |B'(z)| &=& \frac{z^2e^{-A(z)}}{g(z)}
\left|\int^1_z\frac{\Delta T(z')}{z'^2}e^{A(z')}dz'\right|  \nonumber \\
&=&\frac{z^2e^{-A(z)}}{g(z)}
\int^1_z\frac{|\Delta T(z')|}{z'^2}e^{A(z')}dz' \nonumber \\
& \ge & \frac{z^2e^{-A(z)}}{g(z)}e^{A(1)}
\int^1_z\frac{|\Delta T(z')|}{z'^2}dz' \nonumber \\
& \ge & \frac{z^2e^{A(1)-A(1-\epsilon)}}{g(z)}
\int^1_z\frac{|\Delta T(z')|}{z'^2}dz' \,. 
\eea
Here, we used Lemma for the first and second inequalities. 
Substituting Eqs.~(\ref{near-horizon-g}) and (\ref{con-deviation-energy}) into 
(\ref{inequality-B'}), Eq.~(\ref{inequality-B'}) is reduced to 
\bea
\label{sol-B1}
|B'(z)|\ge \frac{u(1)(|\alpha(\kappa)|+O(\epsilon))}{u(1-\epsilon)(\kappa+g_2(\kappa)(1-z))}  \,,
\eea  
for $1-\epsilon\le z\le 1$. 
Note that at the extremal limit $\kappa \to 0$, we can make ${|\alpha(\kappa)|}/{(\kappa+g_2(\kappa)(1-z))}$ as big as $1/\kappa$ as we go near the horizon $z\to 1$. 
This indicates that $B'(z)$ diverges as $1/(1-z)$ 
and $B(z)$ diverges logarithmically near the horizon in the extremal limit. 
One exception for this is, $u(1) \to 0$. However, we have assumed this is not that case by (\ref{hypo1}). 

Since $T_{zz}=R_{zz}\ge 0$ by the condition~1, 
Eq.~(\ref{eq-Raychau1}) means 
\bea
\label{eq-Raychau2}
\theta'\le-\frac{1}{2}B'^2 \,. 
\eea
Therefore, $B'$ diverges as $1/(1-z)$ induces that $\theta$ also diverges 
negatively as $-1/(1-z)$. This indicates that $A$ diverges logarithmically to negative infinity in the extremal limit, therefore $e^A \to 0$, but this contradicts with (\ref{hypo1}). 

To say this in more rigidly, 
using Eq.~(\ref{sol-B1}), we can integrate Eq.~(\ref{eq-Raychau2}) from $u=1-\epsilon$ to 
$u=1$ and thus we obtain 
\bea
\label{eq-theta}
 \theta(z)-\theta(1-\epsilon) \le   
-\frac{u^2(1)(\alpha^2(\kappa)+O(\epsilon))}{2u^2(1-\epsilon)\kappa\, g_2(\kappa)}
\left[\frac{1}{1+\frac{g_2(\kappa)}{\kappa}(1-z)}-\frac{1}
{1+\frac{g_2(\kappa)\epsilon}{\kappa}}
 \right]. 
\eea
Since $\theta(1-\epsilon)\le 0$ by Lemma, it immediately means that 
\bea
\label{eq-theta1}
\theta(z)\le -\frac{u^2(1)(\alpha^2(\kappa)+O(\epsilon))}{2u^2(1-\epsilon)\kappa g_2(\kappa)}
\left[\frac{1}{1+\frac{g_2(\kappa)}{\kappa}(1-z)}-\frac{1}
{1+\frac{g_2(\kappa)\epsilon}{\kappa}}
 \right]. 
\eea
Integrating Eq.~(\ref{eq-theta1}) again from $u=1-\epsilon$ to 
$u=1$, we finally obtain 
\bea
\label{theta-final}
 \frac{u^2(1-\epsilon)}{u^2(1)}\ln \frac{u(1)}{u(1-\epsilon)}\le  
-\frac{\alpha^2(\kappa)+O(\epsilon)}{2g_2(\kappa)}\left[\frac{1}{g_2(\kappa)}
\ln\left(1+\frac{g_2(\kappa)\epsilon}{\kappa}\right)
-\frac{\epsilon}{\kappa+\epsilon g_2(\kappa)}\right] \,. \nonumber \\ 
\eea
This indicates that there is a small positive value $\kappa_0$ such that 
\bea
\frac{u^2(1-\epsilon)}{u^2(1)}\ln \frac{u(1)}{u(1-\epsilon)}\le -D, 
\qquad 0<\kappa\le \kappa_0
\eea
for an arbitrary large positive value $D$. By assumption~(\ref{hypo1}), 
$u(1)\simeq e^{A_0}$ for small positive value $\kappa=\kappa_1<\kappa_0$. 
This would be $u(1-\epsilon;\kappa_1)>1$. This contradicts with the fact that 
$u(1-\epsilon)\le 1$ in Lemma. $\Box$. 

Therefore, at the extremal limit, (\ref{hypo1}) gives the contradiction and 
we conclude that area should approach zero at that limit. 
This is consistent with the numerical solutions we obtained in section 3 and 4, Fig.~1 and 7.


\section{Discussion}
In this paper, 
we have studied several four dimensional gravitational theories 
and we have obtained homogeneous but anisotropic black brane solutions   
given by the metric ansatz 
(\ref{Gaussian-nullMetric1}), which asymptotic to $AdS_4$ space-time. 
The gravitational systems we consider are 1) Einstein-Maxwell dilaton theory, and 
2)  Einstein-Maxwell-dilaton-axion theory with $SL(2,R)$ symmetry.   
The anisotropy is induced in our setting by 
the scalar field profile, either by dilaton $\phi$ or axion $a$. 
The scalar fields have manifest origin of anisotropy and induce the anisotropy for the 
metric while keeping the metric homogeneous at fixed radial slice. 
However, the scalar fields themselves have 
manifest inhomogeneous profile.  
In the case 1), we obtain solutions both analytically and numerically. Analyitic solutions 
are obtained by the perturbation from the isotropic RN black brane solution, and 
in the case 2), we obtain solutions numerically. 
Our solutions are smooth everywhere but seem to approach singular behavior at the horizon in 
the extremal limit, as seen from Fig.~6, 12.

We also showed how the third law of thermodynamics holds in our set-up, which implies that 
as temperature goes to zero, the area also becomes zero. 
Strictly speaking, at the limit temperature goes to zero, 
the solution seems to approach singular behavior. 
As far as we have 
analyzed, we could not obtain 
the regular black brane solution analytically in the extremal limit. 
It is the numerical analysis which suggests that, 
the solution approaches singular in the extremal limit.

Furthermore, even if we do not take the zero temperature limit, 
as we lower the temperature, the curvature of the black brane solutions become bigger 
and bigger near 
the horizon, and there is a critical temperature where two-derivative 
Einstein-Hilbert action breaks down. 
Then,  we have to worry about higher derivative corrections coming from stringy corrections or 
quantum corrections. 
Therefore, precisely speaking, the extremal limit may not be meaningful limit  
in our two derivative Einstein Hilbert action. 
This suggests that we may always need 
non-zero but small temperature as an IR regulator for the system.  

Because we have used the Einstein equations in our proof of third law in section 5, 
our argument at that section is valid for two-derivative Einstein-Hilbert action. 
In holography, this two-derivative assumption in bulk is a big assumption which holds 
only the strong coupling limit in boundary field theory.  
If the coupling constant in field theory is not that big, 
strong coupling expansion in field theory side always induces stringy corrections in the 
gravity side.  
Therefore we should rather regard our argument of third law of thermodynamics as an ``indication'' that as we lower the temperature more, 
the horizon area becomes smaller and smaller. 
It is curious to develop our argument 
furthermore without relying on the two-derivative action assumption.

There are several related questions remaining for such anisotropic solutions. 
One of those is checking the stability. 
It can happen that such anisotropic solutions be unstable by the small perturbations. 
If this is the case, then such solutions give at most meta-stable phases in the dual field theory. 
We left this for future study.

In our paper we have restricted our analysis only to the metric 
of the form (\ref{Gaussian-nullMetric1}), namely we have set $c(z) = 0$ in 
more generic metric (\ref{gaussian-metric}). 
This is because we are considering the 
scalar fields which give $T^y_{\,\,\,\,x} = 0$. 
Obviously with more generic matter contents, we can also obtain anisotropic solutions 
which have nonzero $c(z)$. 
Generic solutions, for examples, as the ones analyzed in 
the Bianchi type \cite{Iizuka:2012iv} does not 
always satisfy $c(z)=0$. Therefore, it is interesting to search more generic 
setting with $c(z) \neq 0$.

As is discussed in section 2, this implies that we need the matter contents which 
satisfy $T^y_{\,\,\,\,x} \neq 0$ in addition to $T^y_{\,\,\,\,y} \neq T^x_{\,\,\,\,x}$. 
For example, if we introduce a scalar field profile such that 
\bea
\label{phifxy}
\phi = f(x, y) \,,
\eea
we can obtain both $T^y_{\,\,\,\,x} \neq 0$ and $T^y_{\,\,\,\,y} \neq T^x_{\,\,\,\,x}$. 
However,  
generically such scalar fields induce inhomogeneous metric, this is 
because  energy-momentum tensor $T^\mu_{\,\,\,\,\nu}$ 
becomes manifestly position-dependent.  
By requiring that an energy-momentum tensor is position-independent, and that it satisfies  
$T^y_{\,\,\,\,x} \neq 0$ and $T^y_{\,\,\,\,y} \neq T^x_{\,\,\,\,x}$, we end up with  
a scalar field profile such that it can depend on coordinate $x$ and $y$ only linearly. 
But this is, after coordinate 
transformation, equivalent to our dilaton profile (\ref{dilatonxdependence}). Therefore 
if we introduce anisotropy by the scalar fields, our setting is quite generic one in that sense.

Of course, we can also induce anisotropy from other degrees of freedom, such as   
the gauge potential \cite{Taylor:2008tg}, \cite{Iizuka:2012iv}, and generical $p$-form potential \cite{Donos:2012gg}. 
Generically, the gauge potentials or $p$-form potentials
which are invariant under Killing vectors respect the isometry of the metric 
and therefore do not induce inhomogeneity. 
Taking the metric and gauge potential made by the one forms of the Bianchi classification,  
we can obtain homogeneous but anisotropy solutions as \cite{Iizuka:2012iv}.
However we can also take the gauge potential, say $A_v$ to be proportional to $x$ 
\bea
\label{Avalphax}
A_v = \alpha x \,,
\eea 
as we took for the scalar fields in this paper. 
In such case, the field strength is position $x$-independent, so is  
the energy-momentum tensor. Therefore, we can obtain homogeneous but anisotropic metric. However, even though we have 
homogeneous metric, the gauge potential is manifestly $x$-dependent and 
we have a source term,  {\it i.e.}, constant flux on the boundary theory. 
We can also construct homogeneous but anisotropic solutions with 
the scalar fields and vector fields where 
scalar fields and gauge potentials are not simply proportional to $x$. By acting the $SL(2,R)$ symmetry 
(\ref{SL(2,R)-transformation1}) and (\ref{SL(2,R)-transformation2}) on the solutions we obtained in section 4, we can have generic solutions where scalar fields and gauge potentials have non-trivial position $x$-dependence. 
Obviously, it is desirable to investigate more generic homogeneous and anisotropic solutions 
where anisotropy is induced not only by the scalars but also by the generic gauge or $p$-form potentials in more detail. 

It would be also interesting if we could construct 
the situation where anisotropy is induced spontaneously as \cite{Donos:2012gg}, instead of the 
situations where anisotropy is induced 
by the source terms (non-normalizable modes) such as (\ref{phifxy}) and (\ref{Avalphax}).  

The metric we consider is such that, the field theory spaces, spanned by $x$ and $y$, 
becomes rectangular, namely the ratio between $x$ and $y$ directions 
changes due to the nonzero $B(z)$. 
This is the situation where one direction is stretched compared with the other. 
For example, if we consider the fermion Green function, then the Fermi-surface should change 
into elliptical shape instead of circle in momentum ($k_x$, $k_y$) space. Since our solutions constructed in section 3 and 4 
have non-trivial gauge potentials $A_v$ which is dual to the chemical potential in boundary theory, 
it is quite interesting to calculate the fermion Green function 
in such anisotropic set-up as \cite{Liu:2009dm,Faulkner:2009wj,Cubrovic:2009ye,Iizuka:2011hg}. 
It is also interesting to study the transport coefficients in such anisotropic settings. 
We left these studies as future projects.

\acknowledgments
We would like to thank N.~Kundu, P.~Narayan, T.~Okamura, N.~Sircar, and S.~P.~Trivedi for discussions. 
K.~M.~is supported in part by the 
Grant-in-Aid for Scientific Research No. 23740200 from the Ministry of Education, 
Culture, Sports, Science and Technology, Japan.

\appendix
\section{Analysis of uniqueness inside the event horizon}
In section 2.2, we have shown that $B=c=0$ in $z\le 1$ in the pure gravity system. This can 
be extended just inside the horizon as follows. Let us suppose that both $B$ and $c$ are 
not zero but small just inside the horizon, $0<z-1 \ll 1$. 
Then, Eqs.~(\ref{eq-B-pg}) and (\ref{eq-C-pg}) are reduced to the following 
linear second order differential equation, 
\bea
\label{eq-regular-singular}
& gK''+g'K'=0,  
\eea
where $K=B,\,c$ and we used the fact that $g$, $c$ are negligible with respect to $g'$, $e^A$ 
in $0<z-1 \ll 1$. 
Since $g\simeq \kappa (1-z)$ near the horizon, Eq.~(\ref{eq-regular-singular}) has a regular 
singular point at $z=1$. Thus, substituting an ansatz $K=(1-z)^\gamma$ into 
Eq.~(\ref{eq-regular-singular}), we obtain $\gamma^2=0$. 
This means that the two independent leading solutions are 
\bea
K_1\simeq 1, \qquad K_2\simeq \ln|1-z|. 
\eea
Using two arbitrary constants $C_1$, $C_2$, the general solution is expressed 
as $K\simeq C_1K_1+C_2K_2$ in $0<z-1 \ll 1$. As $K=B,c=0$ 
at $z=1$, we must take $C_1=C_2=0$. This indicates that anisotropy does not still appear 
even just inside the horizon, i.~e.~, $B,\,c\simeq 0$ when $z-1 \ll 1$.

\section{Scalar curvature invariants for the generalized Gaussian null coordinates}
Given the generalized gaussian null coordinate~(\ref{gaussian-metric}), the scalar curvature 
is calculated as 
\bea
R =\frac{K_R(z)}{2 L^2 \left(e^{2 A(z)}-c(z)^2\right)^2}
\eea
where
\bea
&K_R(z) = 4 z^2 e^{2 A(z)} c(z)^2 g(z) A''(z)-4 z^2 e^{4 A(z)} g(z) A''(z)-8 z^2 e^{2 A(z)} c(z)
   g(z) A'(z) c'(z) \nonumber \\
  & +4 z^2 e^{2 A(z)} c(z)^2 A'(z) g'(z)+7 z^2 e^{2 A(z)} c(z)^2 g(z)
   A'(z)^2-12 z e^{2 A(z)} c(z)^2 g(z) A'(z) \nonumber \\
 &  -4 z^2 e^{4 A(z)} A'(z) g'(z)-3 z^2 e^{4
   A(z)} g(z) A'(z)^2+12 z e^{4 A(z)} g(z) A'(z) 
  \nonumber \\ &  +z^2 e^{2 A(z)} c(z)^2 g(z) B'(z)^2  -z^2
   e^{4 A(z)} g(z) B'(z)^2+4 z^2 e^{2 A(z)} c(z) g(z) c''(z)
   \nonumber \\
 &  +4 z^2 e^{2 A(z)} c(z)
   c'(z) g'(z)+3 z^2 e^{2 A(z)} g(z) c'(z)^2   -12 z e^{2 A(z)} c(z) g(z) c'(z)
   \nonumber \\
 &  +4 z^2 e^{2
   A(z)} c(z)^2 g''(z)  
   -24 z e^{2 A(z)} c(z)^2 g'(z) +48 e^{2 A(z)} c(z)^2 g(z)-2 z^2
   e^{4 A(z)} g''(z)  \nonumber \\
 &   +12 z e^{4 A(z)} g'(z)   -24 e^{4 A(z)} g(z)  -4 z^2 c(z)^3 g(z)
   c''(z)  -4 z^2 c(z)^3 c'(z) g'(z)  \nonumber \\
 & 
   +z^2 c(z)^2 g(z) c'(z)^2+12 z c(z)^3 g(z) c'(z)
   -2 z^2
   c(z)^4 g''(z)   \nonumber \\
 & +12 z c(z)^4 g'(z)-24 c(z)^4 g(z) \,.
\eea
Note that the curvature is non-divergent if $K_R(z)$ is non-divergent and  
\bea
\label{positivedet}
e^{2 A(z)}-c(z)^2 > 0 \,.
\eea
The former condition is satisfied if 
$A$, $A'$, $A''$, $c$, $c'$, $c''$, $g$, $g'$, $g''$, $B'$ are 
regular functions.    
The latter condition implies that 
the determinant of 
the matrix $g_{ij}~(i,j=x,y)$ on the $v=\mbox{const.}$ two dimensional spacelike surface is  
positive (the fixed $v$ slice has positive area). 

Similarly,   
Ricci tensor square $R^{\mu\nu}R_{\mu\nu}$ and 
the Kretschmann scalar curvature invariant $R^{\mu\nu\alpha\beta}R_{\mu\nu\alpha\beta}$ 
are calculated as 
\bea
R^{\mu\nu}R_{\mu\nu} &=& \frac{K_{Ricci}(z)}{L^4 \left(e^{2 A}-c^2\right)^4} \,, \\
 R^{\mu\nu\alpha\beta}R_{\mu\nu\alpha\beta} &=&
\frac{K_{Riemann}(z) }{L^4(e^{2 A}-c^2)^4}, 
\eea
and $K_{Ricci}(z)$ and $K_{Riemann}(z)$ are similar non linear combination of  
$A$, $A'$, $A''$, $c$, $c'$ $c''$, $g$, $g'$, $g''$, $B'$, $B''$,  and non-divergent if these are so. 
Therefore, with (\ref{positivedet})
if the  
functions $A(z)$, $B(z)$, $c(z)$ $g(z)$ are all smooth up to their second derivatives, 
all the scalar curvature invariants 
are finite.



\begin{thebibliography}{99}

\bibitem{Maldacena:1997re} 
  J.~M.~Maldacena,
  ``The Large N limit of superconformal field theories and supergravity,''
  Adv.\ Theor.\ Math.\ Phys.\  {\bf 2}, 231 (1998)
  [Int.\ J.\ Theor.\ Phys.\  {\bf 38}, 1113 (1999)]
  [hep-th/9711200].
  
\bibitem{Witten:1998qj} 
  E.~Witten,
  ``Anti-de Sitter space and holography,''
  Adv.\ Theor.\ Math.\ Phys.\  {\bf 2}, 253 (1998)
  [hep-th/9802150].

\bibitem{Gubser:1998bc} 
  S.~S.~Gubser, I.~R.~Klebanov and A.~M.~Polyakov,
  ``Gauge theory correlators from noncritical string theory,''
  Phys.\ Lett.\ B {\bf 428}, 105 (1998)
  [hep-th/9802109].


\bibitem{Witten:1998zw} 
  E.~Witten,
  ``Anti-de Sitter space, thermal phase transition, and confinement in gauge theories,''
  Adv.\ Theor.\ Math.\ Phys.\  {\bf 2}, 505 (1998)
  [hep-th/9803131].


\bibitem{Gubser:2008px} 
  S.~S.~Gubser,
  ``Breaking an Abelian gauge symmetry near a black hole horizon,''
  Phys.\ Rev.\ D {\bf 78}, 065034 (2008)
  [arXiv:0801.2977 [hep-th]].

\bibitem{Hartnoll:2008vx} 
  S.~A.~Hartnoll, C.~P.~Herzog and G.~T.~Horowitz,
  ``Building a Holographic Superconductor,''
  Phys.\ Rev.\ Lett.\  {\bf 101}, 031601 (2008)
  [arXiv:0803.3295 [hep-th]].


\bibitem{Hartnoll:2008kx} 
  S.~A.~Hartnoll, C.~P.~Herzog and G.~T.~Horowitz,
  ``Holographic Superconductors,''
  JHEP {\bf 0812}, 015 (2008)
  [arXiv:0810.1563 [hep-th]].


\bibitem{Sachdev:2010um} 
  S.~Sachdev,
  ``Holographic metals and the fractionalized Fermi liquid,''
  Phys.\ Rev.\ Lett.\  {\bf 105}, 151602 (2010)
  [arXiv:1006.3794 [hep-th]].



\bibitem{SachdevMueller}
S.~Sachdev and M.~Mueller, 
``Quantum criticality and black holes,'' 
arXiv:0810.3005 [cond-mat.str-el].



\bibitem{Hartnoll:2009sz} 
  S.~A.~Hartnoll,
  ``Lectures on holographic methods for condensed matter physics,''
  Class.\ Quant.\ Grav.\  {\bf 26}, 224002 (2009)
  [arXiv:0903.3246 [hep-th]].
  
\bibitem{Herzog:2009xv} 
  C.~P.~Herzog,
  ``Lectures on Holographic Superfluidity and Superconductivity,''
  J.\ Phys.\ A A {\bf 42}, 343001 (2009)
  [arXiv:0904.1975 [hep-th]].

\bibitem{McGreevy:2009xe} 
  J.~McGreevy,
  ``Holographic duality with a view toward many-body physics,''
  Adv.\ High Energy Phys.\  {\bf 2010}, 723105 (2010)
  [arXiv:0909.0518 [hep-th]].


\bibitem{Horowitz:2010gk} 
  G.~T.~Horowitz,
  ``Introduction to Holographic Superconductors,''
  arXiv:1002.1722 [hep-th].

\bibitem{Sachdev:2010ch} 
  S.~Sachdev,
  ``Condensed Matter and AdS/CFT,''
  arXiv:1002.2947 [hep-th].

\bibitem{CasalderreySolana:2011us} 
  J.~Casalderrey-Solana, H.~Liu, D.~Mateos, K.~Rajagopal and U.~A.~Wiedemann,
  ``Gauge/String Duality, Hot QCD and Heavy Ion Collisions,''
  arXiv:1101.0618 [hep-th].

\bibitem{Hartnoll:2011fn} 
  S.~A.~Hartnoll,
  ``Horizons, holography and condensed matter,''
  arXiv:1106.4324 [hep-th].

 



\bibitem{Kachru:2008yh} 
  S.~Kachru, X.~Liu and M.~Mulligan,
  ``Gravity Duals of Lifshitz-like Fixed Points,''
  Phys.\ Rev.\ D {\bf 78}, 106005 (2008)
  [arXiv:0808.1725 [hep-th]].

\bibitem{Taylor:2008tg} 
  M.~Taylor,
  ``Non-relativistic holography,''
  arXiv:0812.0530 [hep-th].

\bibitem{Li:2009pf} 
  W.~Li, T.~Nishioka and T.~Takayanagi,
  ``Some No-go Theorems for String Duals of Non-relativistic Lifshitz-like Theories,''
  JHEP {\bf 0910}, 015 (2009)
  [arXiv:0908.0363 [hep-th]].

\bibitem{Gubser:2009qt}
  S.~S.~Gubser and F.~D.~Rocha,
  ``Peculiar properties of a charged dilatonic black hole in AdS$_5$,''
  Phys.\ Rev.\ D {\bf 81} (2010) 046001
  [arXiv:0911.2898 [hep-th]].


\bibitem{Goldstein:2009cv} 
  K.~Goldstein, S.~Kachru, S.~Prakash and S.~P.~Trivedi,
  ``Holography of Charged Dilaton Black Holes,''
  JHEP {\bf 1008}, 078 (2010)
  [arXiv:0911.3586 [hep-th]].

\bibitem{Cadoni:2009xm}
  M.~Cadoni, G.~D'Appollonio and P.~Pani,
  ``Phase transitions between Reissner-Nordstrom and dilatonic black holes in 4D AdS spacetime,''
  JHEP {\bf 1003} (2010) 100
  [arXiv:0912.3520 [hep-th]].

\bibitem{Chen:2010kn} 
  C.~-M.~Chen and D.~-W.~Pang,
  ``Holography of Charged Dilaton Black Holes in General Dimensions,''
  JHEP {\bf 1006}, 093 (2010)
  [arXiv:1003.5064 [hep-th]].

\bibitem{Charmousis:2010zz} 
  C.~Charmousis, B.~Gouteraux, B.~S.~Kim, E.~Kiritsis and R.~Meyer,
  ``Effective Holographic Theories for low-temperature condensed matter systems,''
  JHEP {\bf 1011}, 151 (2010)
  [arXiv:1005.4690 [hep-th]].

\bibitem{Perlmutter:2010qu} 
  E.~Perlmutter,
  ``Domain Wall Holography for Finite Temperature Scaling Solutions,''
  JHEP {\bf 1102}, 013 (2011)
  [arXiv:1006.2124 [hep-th]].
  
\bibitem{Bertoldi:2010ca} 
  G.~Bertoldi, B.~A.~Burrington and A.~W.~Peet,
  ``Thermal behavior of charged dilatonic black branes in AdS and UV completions of Lifshitz-like geometries,''
  Phys.\ Rev.\ D {\bf 82}, 106013 (2010)
  [arXiv:1007.1464 [hep-th]].
  
\bibitem{Goldstein:2010aw} 
  K.~Goldstein, N.~Iizuka, S.~Kachru, S.~Prakash, S.~P.~Trivedi and A.~Westphal,
  ``Holography of Dyonic Dilaton Black Branes,''
  JHEP {\bf 1010}, 027 (2010)
  [arXiv:1007.2490 [hep-th]].

\bibitem{Bertoldi:2011zr} 
  G.~Bertoldi, B.~A.~Burrington, A.~W.~Peet and I.~G.~Zadeh,
  ``Lifshitz-like black brane thermodynamics in higher dimensions,''
  Phys.\ Rev.\ D {\bf 83}, 126006 (2011)
  [arXiv:1101.1980 [hep-th]].

\bibitem{Cadoni:2011kv} 
  M.~Cadoni and P.~Pani,
  ``Holography of charged dilatonic black branes at finite temperature,''
  JHEP {\bf 1104}, 049 (2011)
  [arXiv:1102.3820 [hep-th]].

\bibitem{Iizuka:2011hg} 
  N.~Iizuka, N.~Kundu, P.~Narayan and S.~P.~Trivedi,
  ``Holographic Fermi and Non-Fermi Liquids with Transitions in Dilaton Gravity,''
  JHEP {\bf 1201}, 094 (2012)
  [arXiv:1105.1162 [hep-th]].

\bibitem{Berglund:2011cp} 
  P.~Berglund, J.~Bhattacharyya and D.~Mattingly,
  ``Charged Dilatonic AdS Black Branes in Arbitrary Dimensions,''
  arXiv:1107.3096 [hep-th].

\bibitem{Gouteraux:2011qh} 
  B.~Gouteraux, J.~Smolic, M.~Smolic, K.~Skenderis and M.~Taylor,
  ``Holography for Einstein-Maxwell-dilaton theories from generalized dimensional reduction,''
  JHEP {\bf 1201}, 089 (2012)
  [arXiv:1110.2320 [hep-th]].

\bibitem{Iizuka:2012iv} 
  N.~Iizuka, S.~Kachru, N.~Kundu, P.~Narayan, N.~Sircar and S.~P.~Trivedi,
  ``Bianchi Attractors: A Classification of Extremal Black Brane Geometries,''
  arXiv:1201.4861 [hep-th].

\bibitem{Mateos:2011ix} 
  D.~Mateos and D.~Trancanelli,
  ``The anisotropic N=4 super Yang-Mills plasma and its instabilities,''
  Phys.\ Rev.\ Lett.\  {\bf 107}, 101601 (2011)
  [arXiv:1105.3472 [hep-th]].


\bibitem{Mateos:2011tv} 
  D.~Mateos and D.~Trancanelli,
  ``Thermodynamics and Instabilities of a Strongly Coupled Anisotropic Plasma,''
  JHEP {\bf 1107}, 054 (2011)
  [arXiv:1106.1637 [hep-th]].




\bibitem{Nakamura:2009tf} 
  S.~Nakamura, H.~Ooguri and C.~-S.~Park,
  ``Gravity Dual of Spatially Modulated Phase,''
  Phys.\ Rev.\ D {\bf 81}, 044018 (2010)
  [arXiv:0911.0679 [hep-th]];

\bibitem{Ooguri:2010kt} 
  H.~Ooguri and C.~-S.~Park,
  ``Holographic End-Point of Spatially Modulated Phase Transition,''
  Phys.\ Rev.\ D {\bf 82}, 126001 (2010)
  [arXiv:1007.3737 [hep-th]];

\bibitem{Ooguri:2010xs} 
  H.~Ooguri and C.~-S.~Park,
  ``Spatially Modulated Phase in Holographic Quark-Gluon Plasma,''
  Phys.\ Rev.\ Lett.\  {\bf 106}, 061601 (2011)
  [arXiv:1011.4144 [hep-th]].


\bibitem{Donos:2012gg} 
  A.~Donos and J.~P.~Gauntlett,
  ``Helical superconducting black holes,''
  arXiv:1203.0533 [hep-th].





  
  
\bibitem{Shapere:1991ta} 
  A.~D.~Shapere, S.~Trivedi and F.~Wilczek,
  ``Dual dilaton dyons,''
  Mod.\ Phys.\ Lett.\ A {\bf 6}, 2677 (1991).

\bibitem{Bayntun:2010nx} 
  A.~Bayntun, C.~P.~Burgess, B.~P.~Dolan and S.~-S.~Lee,
  ``AdS/QHE: Towards a Holographic Description of Quantum Hall Experiments,''
  New J.\ Phys.\  {\bf 13}, 035012 (2011)
  [arXiv:1008.1917 [hep-th]].

\bibitem{Witten:2003ya} 
  E.~Witten,
  ``SL(2,Z) action on three-dimensional conformal field theories with Abelian symmetry,''
  In *Shifman, M. (ed.) et al.: From fields to strings, vol. 2* 1173-1200
  [hep-th/0307041].



\bibitem{Shapere:1988zv} 
  A.~D.~Shapere and F.~Wilczek,
  ``Selfdual Models with Theta Terms,''
  Nucl.\ Phys.\ B {\bf 320}, 669 (1989).

\bibitem{Lutken:1991jk} 
  C.~A.~Lutken and G.~G.~Ross,
  ``Duality in the quantum Hall system,''
  OUTP-91-19-P.

\bibitem{Kivelson:1992zz} 
  S.~Kivelson, D.~-H.~Lee and S.~-C.~Zhang,
  ``Global phase diagram in the quantum Hall effect,''
  Phys.\ Rev.\ B {\bf 46}, 2223 (1992).

\bibitem{Burgess:2000kj} 
  C.~P.~Burgess and B.~P.~Dolan,
  ``Particle vortex duality and the modular group: Applications to the quantum Hall effect and other 2-D systems,''
  Phys.\ Rev.\ B {\bf 63}, 155309 (2001)
  [hep-th/0010246].
  
\bibitem{dolan:2006zc} 
  B.~P.~Dolan,
  ``Modular Symmetry and Fractional Charges in N=2 Supersymmetric Yang-Mills and the Quantum Hall Effect,''
  SIGMA {\bf 3}, 010 (2007)
  [hep-th/0611282].
 
\bibitem{Liu:2009dm} 
  H.~Liu, J.~McGreevy and D.~Vegh,
  ``Non-Fermi liquids from holography,''
  Phys.\ Rev.\ D {\bf 83}, 065029 (2011)
  [arXiv:0903.2477 [hep-th]].

\bibitem{Faulkner:2009wj} 
  T.~Faulkner, H.~Liu, J.~McGreevy and D.~Vegh,
  ``Emergent quantum criticality, Fermi surfaces, and AdS(2),''
  Phys.\ Rev.\ D {\bf 83}, 125002 (2011)
  [arXiv:0907.2694 [hep-th]].
  
\bibitem{Cubrovic:2009ye} 
  M.~Cubrovic, J.~Zaanen and K.~Schalm,
  ``String Theory, Quantum Phase Transitions and the Emergent Fermi-Liquid,''
  Science {\bf 325}, 439 (2009)
  [arXiv:0904.1993 [hep-th]].

\end{thebibliography}
\end{document}